\documentclass{CSML}

\def\dOi{12(4:11)2016}
\lmcsheading%
{\dOi}
{1--25}
{}
{}
{Nov.~22, 2015}
{Dec.~28, 2016}
{}

\ACMCCS{[{\bf Theory of computation}]: Models of computation---Computability---Lambda calculus; Computational complexity and
  cryptography---Complexity theory and logic; Logic---Proof
  theory\,/\,Constructive mathematics}
\subjclass{[Theory of computation]: 
Proof theory; Constructive mathematics
Lambda-calculus; 
Complexity theory and logic}

\usepackage{hyperref}
\usepackage[utf8]{inputenc}
\usepackage[T1]{fontenc}
\usepackage{amsfonts}\usepackage{amssymb,prooftree}
\usepackage{tikz}\usepackage{xspace}
\newcommand\ciut{\hspace{.3mm}}
\newcommand\nictu{$\phantom{}$}
\def\<{\langle\,}
\def\>{\,\rangle}
\def\To{\Rightarrow}\def\Ot{\Leftarrow}

\def\FV{\mbox{\rm FV\/}}
\def\pusty{\varnothing}
\def\dom#1{{\rm Dom}(#1)}\def\subnoteq{\varsubsetneq}
\def\tam{\rhd} 
\def\T{{\mathcal T}} \def\R{{\mathcal R}}\def\G{{\mathcal G}}
\def\Pos{{\mathcal L}}
\def\NN{\mathbb N} 
 \def\W{{\mathcal W}} \def\Z{{\mathcal Z}} 
\def\Niedelta{\Theta}
\def\wrt{{with respect to\xspace}}
\def\iff{if and only if\xspace}
\newcommand\ok{\mbox{\sc ok}}
\newcommand\start{\mbox{\it start\/}}
\newcommand\lupa{\mbox{\it loop\/}}
\newcommand\pula{\mbox{\it add\/}}

\newcommand\lewy{\mbox{\rm Lt}}
\newcommand\prawy{\mbox{\rm Rt}}
\newcommand\A{{\mathcal A}}\newcommand\M{{\mathcal M}}
\newcommand\F{{\mathcal F}}
\def\I{{\mathcal I}}\def\II{\mathbb I}\def\J{\mathcal J}
\newcommand\rA{{\mathrm A}}%
\newcommand\rE{{\mathrm E}}   
\newcommand\rH{{\mathrm H}}   
   
\newcommand\rT{{\mathrm T}}
\newcommand\rC{{\mathrm C}}
\newcommand\rV{{\mathrm V}}\newcommand\rF{{\mathrm F}}
\newcommand\rS{{\mathrm S}}
\newcommand\rL{{\mathrm L}}\newcommand\rB{{\mathrm B}}
\newcommand\rK{{\mathrm K}}
\newcommand\rM{{\mathrm M}}\newcommand\rN{{\mathrm N}}
\newcommand\rj{{\mathbf 1}}\newcommand\rd{{\mathbf 2}}
\newcommand\rp{{\mathbf p}}\newcommand\rP{{\mathrm P}}
\newcommand\rqq{{\mathbf q}}\newcommand\rQ{{\mathrm Q}}
\newcommand\rR{{\mathrm R}}
\newcommand\rU{{\mathrm U}}\newcommand\rX{{\mathrm X}}
\newcommand\bus{\mathrm{Bus}}

\newcommand\TG{\G^{*}\xspace}
\theoremstyle{plain}
\newtheorem{example}[thm]{Example}
\newtheorem{proposition}[thm]{Proposition}
\newcount\hour \hour=\time
\newcount\minute \minute=\hour
\divide\hour by 60
\newcount\temp \temp=\hour
\multiply\temp by 60
\advance\minute by -\temp
\newcount\temp \temp=\minute
\divide\temp by 10
\def\optzero{\ifcase\temp 0\fi}

\def\miesiaca{\ifcase\month\or stycznia\or lutego\or marca\or kwietnia%
\or maja%
\or czerwca\or lipca\or sierpnia\or wrze\'snia%
                           \or pa\'zdziernika\or listopada\or grudnia\fi}

\begin{document}

\title[On the Mints Hierarchy]%
{On the Mints Hierarchy in First-Order \\
Intuitionistic Logic\rsuper*} 
\titlecomment{{\lsuper*}Project supported through NCN 
grant~DEC-2012/07/B/ST6/01532. This paper is a revised and expanded 
version of~\cite{suz-fos15}.}

\author[A.~Schubert]
{Aleksy Schubert\rsuper a}	
\address{{\lsuper{a,b}}Institute of Informatics, University of Warsaw\\
  ul. S. Banacha 2, 02--097 Warsaw, Poland}	
\email{\{alx, urzy\}@mimuw.edu.pl}  

\author[P.~Urzyczyn]{Paweł Urzyczyn\rsuper b}	
\address{\vspace{-18 pt}}	

\author[K.~Zdanowski]{Konrad Zdanowski\rsuper c}	
\address{{\lsuper c}Cardinal Stefan Wyszyński University in Warsaw\\
  ul. Dewajtis 5, 01-815 Warsaw, Poland}	
\email{k.zdanowski@uksw.edu.pl}  





\begin{abstract}
We stratify intuitionistic first-order logic over
$(\forall,\to)$ into fragments 
  determined by the alternation of positive
  and negative occurrences of quantifiers (Mints hierarchy).
  We study the decidability and complexity of these fragments. 
We prove that even the $\Delta_2$ level is undecidable 
and that $\Sigma_1$ is {\sc Expspace}-complete. We also prove that
the arity-bounded  fragment of $\Sigma_1$ is complete for
{\it co}-{\sc Nexptime}.
\end{abstract}
\maketitle

\section{Introduction}
\label{sec:introduction}

\noindent
The leading proof assistants such as Coq \cite{Coq}, Agda \cite{Agda}
or Isabelle~\cite{Paulson89} are founded on constructive logics. 
Still, the complexity behind proof search in constructive reasoning
systems is not well understood even for their basic and crucial fragments where
the implication and universal quantification are used. This situation
is caused partly by the difficulty of the field and partly by the lack
of a systematic approach, especially in the case of quantifiers.

Quantifiers are present in logic at least from the time of Aristotle but
a~modern theory of quantification was probably initiated by 
Ch.S.~Peirce~and G.~Frege~\cite{Bonevac}.
The  systematic approach to quantifiers through their grouping at the 
beginning of a logical formula  was originated by Peirce 
and worked out by A.~Church~\cite{church-iml},
 who first used the term ``prenex normal form''. Since then classifying 
formulas according
to the quantifier prefix remains a~standard stratification tool 
in modern logic, just to mention Ehrenfeucht-Fraïssé 
games \cite[Chapter~6]{Immerman99} or 
the arithmetical hierarchy of Kleene and 
Mostowski~\cite[Chapter~7]{Fitting81}.

Classes of prenex formulas in the full first-order language,
beginning with $\exists$ (resp.~$\forall$), and 
with $n$ alternating groups of quantifiers are denoted
in this paper by the sans-serif symbol~${\sf\Sigma}_n$ (resp.~${\sf\Pi}_n$).
(The ordinary serifed symbols $\Sigma$ and $\Pi$ are reserved for 
classes of the Mints hierarchy.) 
It is known that classes ${\sf\Sigma}_n$ and ${\sf\Pi}_n$ form
a strict hierarchy \wrt~their classical expressive power~\cite{rosen05}.
While the prenex normal form is useful for classification
of formulas, which was demonstrated in full strength by
Börger, Grädel, and Gurevich in their influential book
\cite{BorgerGG1997}, it is rarely used in practice.
The structure of formulas
arising from actual reasoning (in particular proof formalization) 
often  involves quantification in arbitrary positions. For instance 
this happens when a~quantified definition is expanded in a formula.

In addition, the prenex normal form theorem applies to 
{\it classical\/} logic only. Things become
quite different for constructive logic (aka intuitionistic logic),
because the prenex fragment of intuitionistic logic is
decidable~\cite{rash54}. This contrasts with the undecidability of
the general case (see e.g.,~\cite{sorm06})
and that makes this form of stratification
unsuitable in the constructive context.

Can we replace the prenex classification by something adequate 
for intuitionistic logic? Yes, we can: as observed by 
Grigori Mints~\cite{minc},  the 
principal issue is the alternation of positive
and negative occurrences of quantifiers in a~formula, 
understood as in~\cite{Wang60}. Roughly speaking, a quantifier occurrence
is positive iff a classical reduction to a prenex form turns it
into a universal quantifier. Dually, negative occurrences of quantifiers
are those which become existential quantifiers after normalization. 
This yields the {\it Mints hierarchy\/} of formulas, consisting of the
following classes (note the serifed $\Sigma$ and $\Pi$): 
\begin{itemize}[label=$\Pi_1$]
\item[$\Pi_1$] -- All quantifier occurrences are positive.
\item[$\Sigma_1$] -- All quantifiers  occurrences are  negative.
\item[$\Pi_2$] -- Up to one alternation: no positive 
quantifier in scope of a negative one.
\item[$\Sigma_2$] -- Up to one alternation: no negative 
quantifier in scope of a positive one.
\end{itemize}
And so on. Every  
formula can be classified as a $\Pi_n$ or a $\Sigma_n$ formula
without actually reducing it to a prenex form. Therefore, Mints hierarchy 
makes perfect sense for intuitionistic logic. 

In this paper we address the question of decidability and complexity 
of the intuitionistic provability problem for classes $\Pi_n$ and $\Sigma_n$. 
This of course resembles the subject of~\cite{BorgerGG1997}, and it is 
natural to compare our results with those in the book. As it may be
expected, the intuitionistic case is 
at least as hard as 
the classical case. (Remember though that complexity results 
about classical logic are usually stated in terms of satisfiability.)

As for the existing knowledge, Mints
proved that the fragment $\Pi_1$ of the constructive logic with all
connectives and quantifiers is decidable~\cite{minc}.  
An alternative proof of Mints' 
result (for the calculus with $\forall$ and $\to$ only) was given by
Dowek and Jiang \cite{dowekjiang06}. A~similar 
decidability result was also obtained by Rummelhoff~\cite{rummelhoff}
for the positive fragment of second-order propositional intuitionistic
logic (system~F). The  {\it co-}2-{\sc Nexptime} lower bound for $\Pi_1$
was proved by Schubert, 
Urzyczyn and Walukiewicz-Chrząszcz \cite{suw2013}, but the
problem is conjectured
to be non-elementary~\cite{suw2013-www}. 
The undecidability of $\Sigma_2$ with all connectives and quantifiers
can be derived from the undecidability of the classical satisfiability 
problem for~$\forall^*\exists^*$ using a result of 
Kreisel~\cite[Thm.~7]{Kreisel58}. This would not work for~$\Pi_2$
because the classical satisfiablity of the Ramsey class 
$\exists^*\forall^*$ 
is decidable. Undecidability for $\Pi_2$ (for the full language with 
one binary predicate) is implied by a result of Orevkov~\cite{orewkow65}.
The conference version~\cite{suz-fos15} of the present paper 
strenghtened Orevkov result by showing the undecidability
for the $(\forall,\to)$-fragment.

There are other forms of quantifier-oriented 
hierarchical stratifications of intuitionistic formulas.
For instance, the classical prenex hierarchy can be embedded in
a fragment of the intuitionistic logic: a negation of a prenex formula 
is classically 
provable \iff~it is provable intuitionistically~\cite{Kreisel58}.
A similar, but more general class of formulas in so
called pseudoprenex form, where quantifiers may be separated by double
negation $\lnot\lnot$, was studied in depth by Orevkov 
who gave a full characterization 
 of decidable cases \cite{orewkow76}. 
Also a full characterization of decidable cases was given for prenex 
formulas with 
equality and function symbols~\cite{DegtyarevGNVV00}. Other hierarchies 
of intuitionistic formulas were proposed e.g.,~by 
Fleischmann~\cite{Fleischmann10} and Burr~\cite{Burr09}
(the latter for arithmetic). However, we are not aware of 
any complexity-oriented results for those hierarchies.

In this paper we expand the systematic study of the decision
problem in Mints hierarchy initiated in~\cite{suz-fos15}. Basically, 
we restrict attention to the 
fragment where only the implication and the universal 
quantifier may occur.
Our main results are as follows:
\begin{itemize}[label=\Alph*.]
\item[A.] The hierarchy is strict \wrt~the expressive power.
\item[B.] The decision problem for the class $\Delta_2=\Sigma_2\cap\Pi_2$ 
 is undecidable.
\item[C.] The decision problem for the class $\Sigma_1$ is 
{\sc Expspace}-complete.
\item[D.] The decision problem for arity-bounded $\Sigma_1$ formulas is 
{\it co}-{\sc Nexptime} complete.
\item[E.] The decision problem for $\Sigma_1$ restricted to any 
finite signature is in {\it co}-{\sc Nexptime}.
\end{itemize}
These results are supplemented by the {\it co-}2-{\sc Nexptime}
lower bound for $\Pi_1$
obtained in~\cite{suw2013} and a strong evidence towards the conjecture
that $\Pi_1$ is actually non-elementary~\cite{suw2013-www}. 
Observe that, because of conservativity, part B
applies directly to the full intuitionistic logic, and the same holds for 
the lower bound in~C. The upper bound in C also extends to the general case
at the cost of some additional complication. 

The undecidabilities in B
are shown for the monadic fragment of minimal logic 
(i.e.,~the language with only unary predicate symbols). 
Our proof of A requires a binary predicate 
but we conjecture that the monadic hierarchy is also strict.
It is slightly different with C versus~D, where we have arrived at 
the open problem whether {\it co}-{\sc Nexptime} equals {\sc Expspace}.

The paper is organized as follows. Section~\ref{sec:preliminaries}
contains the basic definitions, and proves strictness of the 
hierarchy. 
Section~\ref{sec:machines} introduces the undecidable tiling puzzles.
Those are encoded in Section~\ref{sec:sigmapidwa} into $\Delta_2$ 
formulas. In Section~\ref{subsec:mon} we use a syntactic translation
to obtain undecidability  for the monadic fragment
of~$\Delta_2$. In Section~\ref{sec:sigmajeden}
we show {\sc Expspace}-completeness for $\Sigma_1$ using the 
decision problem for bus machines~\cite{urzy09}. 
In~the last Section~\ref{sec:jhuyhjy} we study 
$\Sigma_1$ formulas with predicates of bounded arity.

\section{Preliminaries}
\label{sec:preliminaries}

\noindent
We consider first-order intuitionistic logic without function symbols 
and without equality. That is, the only individual
terms are {\it object variables\/}, written in lower case, e.g., $x,y,\dots$
In this paper we restrict 
attention to formulas built only from implication and the universal
quantifier. 
A formula is therefore either an atom $\rP(x_1,\ldots,x_n)$, where $n\geq 0$,
or an implication $\varphi\to\psi$, or it has the form $\forall x\,\varphi$.

We use common parentheses-avoiding  conventions, in particular 
we take the implication to be right-associative. That is,
$\varphi\to\psi\to\vartheta$ stands for $\varphi\to(\psi\to\vartheta)$.

Our proof notation is an extended lambda-calculus of 
{\it proof terms\/} or simply {\it proofs\/} or 
{\it terms\/}. Formulas are treated as \relax
types assigned to proof terms. 
In addition to object variables, in proof terms there are also 
{\it proof variables\/}, written as upper-case letters, like $X$,
$Y$, $Z$. 
  An~{\it environment\/} is a set of declarations
  $(X:\varphi)$, where~$X$ is a~proof variable and~$\varphi$ is
  a~formula.
The type-assignment rules in Figure~\ref{prufassirules} infer 
judgments of the form $\Gamma \vdash M: \varphi$, where~$\Gamma$ 
is an~environment, $M$ is a~proof term, and $\varphi$ is a formula.
In $({\forall}{I})$ we require $x \not \in \FV(\Gamma)$
and $y$ in~$({\forall}{E})$ is an arbitrary object variable.
\begin{figure}[h]
$$
\Gamma, X: \varphi ~\vdash X:\varphi \quad({A}x)
$$
$$\begin{prooftree}
\Gamma, X: \varphi ~\vdash M:\psi
\justifies
\Gamma ~\vdash \lambda X\ciut{:}\ciut\varphi.\ciut M\ :\ \varphi \to \psi
\thickness=0.08em
\using
	({\to}{I})
\end{prooftree}
\qquad
\begin{prooftree}
\Gamma ~\vdash M : \varphi \to \psi \quad
\Gamma ~\vdash N : \varphi
\justifies
\Gamma ~\vdash MN : \psi
\thickness=0.08em
\using
	({\to}{E})
\end{prooftree}$$
$$\begin{prooftree}
\Gamma ~\vdash M : \varphi
\justifies
\Gamma ~\vdash \lambda x\,M : \forall x \varphi
\thickness=0.08em
\using
	({\forall}{I})
\end{prooftree}
\qquad
\begin{prooftree}
\Gamma ~\vdash M : \forall x \varphi
\justifies
\Gamma ~\vdash My : \varphi[x:=y]
\thickness=0.08em
\using
	({\forall}{E})
\end{prooftree}$$~
\caption{Proof assignment rules\label{prufassirules}}
\end{figure}

\noindent That is, we have two kinds of lambda-abstraction: the 
proof abstraction \mbox{$\lambda X\ciut{:}\,\varphi.\,M$} and
the object abstraction $\lambda x\,M$.
There are also two forms of application: the proof application~$MN$, where $N$
is a proof term, and the object application~$My$, where~$y$ is an object
variable. We use the conventions common in lambda-calculus 
e.g., unnecessary parentheses are omitted
and the application is left-associative: $MNP$ means $((MN)P)$.
 Terms and formulas 
are taken up to alpha-conversion.

The formalism is used liberally. Terms are always assumed to be well-typed,
even if type information is left out. \relax
For instance, we often say 
that ``a term $M$ has type $\varphi$'' leaving
the environment implicit. 
Also we often identify environments with
sets of formulas, as well as we write $\Gamma\vdash\varphi$ 
when $\Gamma\vdash M:\varphi$ and $M$ is not relevant at the moment. 
 Sometimes for convenience we drop~$\varphi$ from $\lambda
  X\,{:}\,\varphi.\,M$ when it 
can be deduced from the context. 

Free (object) variables $\FV(\varphi)$ in a formula $\varphi$ are 
as usual. We also define free variables in proofs:
\mbox{$\FV(X)=\pusty$}, $\FV(\lambda X\,{:}\,\varphi.\, M)=
\FV(\varphi)\cup\FV(M)$, \mbox{$\FV(MN)=\FV(M)\cup\FV(N)$}, $\FV(\lambda x\,M)=
\FV(M)-\{x\}$, $\FV(My)=\FV(M)\cup\{y\}$.
The notation
$M[\vec x:=\vec y\ciut]$ stands for the simultaneous substitution of 
a vector of \relax
variables $\vec y=y_1\ldots y_n$ for free occurrences of (different) 
variables $\vec x=x_1\ldots x_n$. 
To make this precise, we take:
\begin{itemize}
\item $x_i[\vec x := \vec y\ciut]=y_i$, and $z[\vec x := \vec y\ciut]=z$, when 
$z$ is not in $\vec x$; 
\item $X[\vec x := \vec y\ciut]=X$;
\item $(\lambda X\,{:}\,\varphi.\, M)[\vec x := \vec y\ciut]=
  \lambda X\,{:}\,\varphi[\vec x := \vec y\ciut].\, M[\vec x := \vec y\ciut]$,
  where $\varphi[\vec x := \vec y\ciut]$ is as usual;
\item 
$(MN)[\vec x := \vec y\ciut]=M[\vec x := \vec y\ciut]N[\vec x := \vec y\ciut]$;
\item 
$(\lambda x\,M)[\vec x := \vec y\ciut]=\lambda x\,M[\vec x := \vec y\ciut]$,
when $x$ is not among $\vec x,\vec y$;
\item 
$(My)[\vec x := \vec y\ciut]=M[\vec x := \vec y\ciut]y[\vec x := \vec y\ciut]$.
\end{itemize}

\begin{lem}\label{lkiojiul}
If $\Gamma\vdash N:\varphi$ then 
$\Gamma[\vec x := \vec y\ciut]\vdash N[\vec x := \vec y\ciut]:
\varphi[\vec x := \vec y\ciut]$.
\end{lem}
\proof  Easy induction.\qed

A term is in {\it normal form\/} when it 
contains no redex, i.e., no subterm of the form
$(\lambda X\,{:}\,\varphi.\, M)N$ or of the form 
$(\lambda x\, M)y$. We also define the 
notion of a proof term in {\it long normal form\/}, abbreviated~{\it lnf\/}.
\begin{itemize}
\item If $N$ is an lnf of type $\varphi$ then $\lambda x\, N$ is an lnf of
  type $\forall x\,\varphi$.\relax
\item If $N$ is an lnf of type $\psi$ then 
$\lambda X\,{:}\,\varphi.\, N$  is an lnf 
of type $\varphi \to \psi$.
\item If $N_1,\ldots, N_n$ are lnf or object variables, and $XN_1\ldots
  N_n$ is of an atom type, then $XN_1\ldots N_n$ is an lnf.
\end{itemize}
The following lemma is shown in~\cite{suw2013-www}.

\begin{lem} 
\label{lemma:lnf}
If $\varphi$ is intuitionistically derivable from $\Gamma$ then \relax
$\Gamma \vdash N:\varphi$, for some lnf~$N$.\qed
\end{lem}

\noindent
The {\it target\/} of a formula is the relation symbol at the end of it.
Formally, 
\mbox{${\it target\/}(\rP(\vec x))=\rP$}, for atomic formulas,
${\it target\/}(\varphi\to\psi)={\it target\/}(\psi)$, 
and ${\it target\/}(\forall x\,\varphi)={\it target\/}(\varphi)$. 
The following observation is essential in long normal proof search.
\begin{lem}\label{lemma:lnfgggrd}
If $\Gamma\vdash N:\rP(\vec x)$, where $\rP(\vec x)$ is an atomic formula
and $N$ is an lnf, then $N=X\vec D$, where $(X:\psi)\in\Gamma$ with
${\it target\/}(\psi)=\rP$, and $\vec D$ is a sequence that may contain
proof terms and object variables. 
\end{lem}\proof  An easy consequence of the definition of an lnf.
\relax\qed

\paragraph{Miscellaneous:} The set of all words over an alphabet $\A$
 is written as $\A^*$. By $\varepsilon$ we denote the empty word.
The relation $w\subseteq v$ holds when $w$ 
is a prefix of~$v$.

\subsection{An example}

To illustrate the computational flavour of intuitionistic proof search 
we consider the formula 
$\alpha_0\to\alpha_1\to\alpha_2\to\alpha_3\to\beta\to\rC$, where $\rC$ 
is a nullary atom and:
\begin{itemize}[label=$\ \alpha_0{=}$]
\item[$\alpha_0=$] $\forall x(\rF(0,x)\,{\to}\,\rS(0,x)\,{\to}\,
\rT(0,x)\,{\to}\, \lupa(x))\,{\to}\, \rC$;
\item[$\alpha_1=$] $\forall x(\rF(0,x)\,{\to}\, 
                     \forall y(\rF(1,y)\,{\to}\,
                      \forall z(\rS(z,x)\,{\to}\,\rS(z,y))\,{\to}\,$\\
\nictu\hspace{4.45cm}$\forall z(\rT(z,x)\,{\to}\,\rT(z,y))\,{\to}\, 
                                                 \lupa(y))\,{\to}\,\lupa(x))$;
\item[$\alpha_2=$] $\forall x(\rF(1,x)\,{\to}\, \rS(0,x)\,{\to}\, 
                     \forall y(\rF(0,y)\,{\to}\, \rS(1,y)\,{\to}\,$\\
\nictu\hspace{4.45cm}$\forall z(\rT(z,x)\,{\to}\,\rT(z,y))\,{\to}\, 
                                              \lupa(y))\,{\to}\, \lupa(x))$;
\item[$\alpha_3=$] 
$\forall x(\rF(1,x)\,{\to}\, \rS(1,x)\,{\to}\, \rT(0,x)\,{\to}\,$\\
\nictu\hspace{4.45cm}
$\forall y(\rF(0,y)\,{\to}\, \rS(0,y)\,{\to}\, \rT(1,y)\,\to 
                                  \lupa(y))\,{\to}\, \lupa(x))$;

\item[$\beta=$] $\forall x(\rF(1,x)\to\rS(1,x) \to \rT(1,x)\to\lupa(x))$.
\end{itemize}\bigskip
In the above, $0$ and $1$ are fixed free variables, playing the role 
of ``bits''. The predicates $\rF$(irst), $\rS$(econd), and $\rT$(hird),
are intended to always occur together to associate three bits to a~variable.
For instance, assumptions $\rF(1,x), \rS(0,x), \rT(1,x)$ associate 
the binary string $101$ to the variable~$x$.
Our formula is constructed 
in such a way that every proof of it must ``generate'' variables associated
with all binary strings of length three. 

To derive $\rC$ from $\Gamma=\{\alpha_0,\alpha_1,\alpha_2,\alpha_3,\beta\}$, 
one must use an assumption with target $\rC$, and~$\alpha_0$
is the only such assumption. So we need to prove 

\hfil $\Gamma\vdash \forall x(\rF(0,x)\,{\to}\,\rS(0,x)\,{\to}\,
\rT(0,x)\,{\to}\, \lupa(x))$, 

\noindent and this amounts to proving 
$\Gamma\vdash \rF(0,x_1)\,{\to}\,\rS(0,x_1)\,{\to}\,
\rT(0,x_1)\,{\to}\, \lupa(x_1)$, where $x_1$ is a~fresh eigenvariable. 
That is, we now have the new proof goal $\lupa(x_1)$
to be derived using additional assumptions 
$\rF(0,x_1),\rS(0,x_1),\rT(0,x_1)$. (We interpret it as ``$x_1$ is associated
to the string $000$''.) 
Given this knowledge about $x_1$
we readily discover that $\alpha_1$ is the only applicable assumption,
as otherwise we would have to prove $\rF(1,x_1)$, which is clearly hopeless.
So we instantiate $\alpha_1$ with $x_1$ in place of $x$ and we now need
to derive the universal formula $\forall y(\rF(1,y)\,{\to}\,
\forall z(\rS(z,x_1)\,{\to}\,\rS(z,y))\,{\to}\,
\forall z(\rT(z,x_1)\,{\to}\,\rT(z,y))\,{\to}\, \lupa(y))$. 
This introduces a~new eigenvariable $x_2$. Our new goal is $\lupa(x_2)$,
and our new assumptions are $\rF(1,x_2)$ and 
$\forall z(\rS(z,x_1)\,{\to}\,\rS(z,x_2))$,
$\forall z(\rT(z,x_1)\,{\to}\,\rT(z,x_2))$. The latter two 
can be used (if~needed) to derive $\rS(0,x_2)$ and $\rT(0,x_2)$.
(This implicitly assigns the string $100$ to $x_2$.)
With this knowledge at hand, we can now try to apply $\alpha_2$
towards proving~$\lupa(x_2)$. We leave it to the reader to check that
our proof construction will lead us to introducing (at least) six 
other eigenvariables
$x_3,\ldots,x_8$ and that the proof will be completed with an
application of $\beta$, when we reach the string 111, i.e., when the assumptions
$\rF(1,x_8),\rS(1,x_8),\rT(1,x_8)$ become available. Note that various
instances of $\alpha_1$ occur in the proof four times, and $\alpha_2$
is used twice.

If we use proof variables $X_0,X_1,X_2,X_3,Y$ to denote assumptions 
$\alpha_0,\alpha_1,\alpha_2,\alpha_3,\beta$, respectively, then the proof can
be written as the following lambda-term. The 
possibly confusing subterm $Z^4_30(Z^3_30(Z^2_30Z^1_3))$ has type 
$T(0,x_4)$ and corresponds to a composition of
assumptions \mbox{$Z^{i+1}_3:\forall z(\rT(z,x_i)\,{\to}\,\rT(z,x_{i+1}))$}, 
for $i=1,2,3$, applied to the assumption $Z_3^1:\rT(0,x_1)$.\medskip

\hspace{1.25cm}
$\lambda X_0X_1X_2X_3Y.\,
X_0(\lambda x_1\lambda Z^1_1Z^1_2Z^1_3.\,\\\nictu\hspace{4.15cm}
X_1x_1Z^1_1(\lambda x_2\lambda Z^2_1Z^2_2Z^2_3.\,\\\nictu\hspace{4.15cm}
X_2x_2Z^2_1(Z^2_20Z^1_2)(\lambda x_3\lambda Z^3_1Z^3_2Z^3_3.\,\\
\nictu\hspace{4.15cm}
X_1x_3Z^3_1(\lambda x_4\lambda Z^4_1Z^4_2Z^4_3.\,\\\nictu\hspace{4.15cm}
X_3x_4Z^4_1(Z^4_21Z^3_2)(Z^4_30(Z^3_30(Z^2_30Z^1_3))) 
(\lambda x_5\lambda Z^5_1Z^5_2Z^5_3.\,\\\nictu\hspace{4.15cm}
X_1x_5Z^5_1%
(\lambda x_6\lambda Z^6_1Z^6_2Z^6_3.\,\\
\nictu\hspace{4.15cm}
X_2x_6Z^6_1(Z^6_20Z^5_2)(\lambda x_7\lambda Z^7_1Z^7_2Z^7_3.\,\\
\nictu\hspace{4.15cm}
X_1x_7Z^7_1(\lambda x_8\lambda Z^8_1Z^8_2Z^8_3.\,\\
\nictu\hspace{4.15cm}
Yx_8Z^8_1(Z^8_21Z^7_2)(Z^8_31(Z^7_31(Z^6_31Z^5_3))))))))))) 
$\vspace{2mm}

\noindent
The above proof is the shortest normal proof of our formula. Other proofs 
may ``generate'' additional variables associated to various strings.
However, repeated strings are ``redundant'', i.e.,  they 
do not help to  complete the proof.

\subsection{The Mints hierarchy}\label{theminc}

We define classes of formulas $\Sigma_n$ and $\Pi_n$ by induction,
beginning with $\Sigma_0=\Pi_0$ being the set of quantifier-free formulas.
The induction step can be expressed by the following 
pseudo-grammar: 
\begin{itemize}
\item $\Sigma_{n+1} ::= {\bf a}\ |\ \Pi_{n}\ |\ \Pi_{n+1}\to \Sigma_{n+1}$
\item $\Pi_{n+1} ::= {\bf a}\ |\ \Sigma_n\ |\ \Sigma_{n+1}\to\Pi_{n+1}\ |\ 
\forall x\;\Pi_{n+1}$
\end{itemize}
where the metavariable {\bf a} stands for an atom.
In addition, we take:
\begin{itemize}
\item $\Delta_n = \Sigma_{n}\cap \Pi_n$.
\end{itemize}

\noindent For example, the formula $(\forall x\,\rP(x)\,{\to}\,\rQ)\to \rQ$ is
in $\Pi_1$, the formula 
\mbox{$\forall x(\forall y\,\rR(y)\,{\to}\,\rP(x))\to \rQ$}
is in $\Sigma_2$, and 
$\forall x\,\rP(x)\to (\forall y\,\rR(y) \to  \rQ)\to \rQ$ is in~$\Delta_2$.
By an easy induction one proves that every $\Sigma_{n}$ formula 
is classically equivalent to a prenex formula of type~${\sf\Sigma}_n$
(recall that the sans-serif ${\sf\Sigma}$ and ${\sf\Pi}$ refer to the ordinary 
classical hierarchy of prenex forms), 
and similarly for $\Pi_n$ versus~${\sf\Pi}_n$. The converse is not true
in the following sense. Consider the formula 
\mbox{$\varphi=\forall x(((\rP(x)\to\rR(x))\to\rP(x))\to\rP(x))$}.
As a~classical tautology, $\varphi$~is classically equivalent 
to an arbitrary quantifier-free tautology, i.e., it is classically 
equivalent to a formula in~${\sf\Sigma_0}$. But in the intuitionistic
logic $\varphi$ is not equivalent to any open ($\Sigma_0$) formula. 

To prove that the Mints hierarchy is strict, we use an analogous 
result about classical logic. 
The following theorem follows from \cite{rosen05}.

\begin{thm}[Rosen]\label{Rosen-thm}
For each $n$ there is a ${\sf\Sigma}_n$ formula $\varphi_n$ which 
is not classically equivalent to any ${\sf\Pi}_n$ formula 
and there is a ${\sf\Pi}_n$ formula $\psi_n$ such that 
$\psi_n$ is not classically equivalent
to any ${\sf\Sigma}_n$ formula. Both formulas are in a language with one binary
predicate.\qed
\end{thm}

Since conjunction and disjunction 
are classically definable from $\to$ and $\bot$, it follows that 
Theorem~\ref{Rosen-thm} holds for the language with $\to$ and $\bot$ 
as  the only propositional connectives. 
If we replace all existential quantifiers in a ${\sf\Sigma}_n$
(resp.~${\sf\Pi}_n$) formula
by their classical definitions in terms of $\forall$, $\to$ and $\bot$, we
obtain a formula which is {\it almost\/} a $\Sigma_n$ (resp.~$\Pi_n$) 
formula in our sense.
Since intuitionistic provability implies classical provability, 
the formula $\varphi_n$ in Theorem~\ref{Rosen-thm} must not be 
intuitionistically
equivalent to any $\Pi_n$ formula. This immediately implies a hierarchy 
theorem for intuitionistic logic with $\forall$, $\to$ and $\bot$ and one 
binary predicate.
To get rid of $\bot$, we use the following obvious lemma, where 
$\vdash_c$ refers to classical provability. 

\begin{lem}\label{AnObviousLemma}
Let $\Gamma\vdash_c \varphi$, and let $p$ be a nullary relation symbol. 
Then $$\Gamma[p:=\bot]\vdash_c \varphi[p:=\bot].$$
\end{lem}
\proof  Routine induction. 
\qed

\begin{cor}\label{Rosen-nobot}
For each $n$ there is a $\Sigma_n$ formula $\varphi_n$ which 
is not classically equivalent to any $\Pi_n$ formula 
and there is a $\Pi_n$ formula $\psi_n$ which 
is not classically equivalent to any $\Sigma_n$ formula. The 
formulas $\varphi_n$ and $\psi_n$ are in a language with 
one binary and one nullary predicate. 
\end{cor}
\proof 
Let $\varphi_n'$ and $\psi_n'$ be the formulas from Theorem~\ref{Rosen-thm},
and let $\varphi_n''$ and $\psi_n''$ be obtained respectively from 
$\varphi_n'$ and $\psi_n'$,
by replacing each $\exists$ by $\neg\forall\neg$ and then eliminating
the connectives $\vee$, $\wedge$, and $\neg$ in a standard way. Finally,
we replace all occurrences of $\bot$ in $\varphi_n''$ and~$\psi_n''$
with a~new nullary predicate symbol, say $p$, 
and we denote the results by $\varphi_n$ and $\psi_n$. 
Let~$\vartheta$ be a~$\Pi_n$ formula. 
If $\varphi_n$ and $\vartheta$ are classically equivalent 
then by Lemma~\ref{AnObviousLemma} so are the formulas $\varphi_n''$ 
and $\vartheta[p:=\bot]$. 
Hence $\varphi_n'$ is classically equivalent to a~${\sf\Sigma}_n$ 
formula, contradicting Theorem~\ref{Rosen-thm}. A~similar argument 
applies to~$\psi_n$. 
\qed

Now, as an easy consequence we can state the following.

\begin{thm}
The Mints hierarchy is strict, that is,
for each $n$ there is a $\Sigma_n$ formula $\varphi_n$ which 
is not intuitionistically  equivalent to any $\Pi_n$ formula 
and there is a $\Pi_n$ formula $\psi_n$ which
is not intuitionistically equivalent to any $\Sigma_n$ formula. 
The formulas $\varphi_n$ and $\psi_n$ are in a language with 
one binary and one nullary predicate. \qed
\end{thm}

\section{Machines and tilings}
\label{sec:machines}

\noindent 
To give a concise account of our lower bound results, we disguise  
 Turing Machines as tiling problems, cf.~\cite[Chapter 3.1.1]{BorgerGG1997}.
While the masquerade is quite obvious to unveil, 
it is still useful: some formulas become simpler. 
In the following two subsections,  we define two forms of slightly
unusual tiling puzzles. Deterministic puzzles of Section~\ref{sec:procedures}
are later used for the undecidability of $\Delta_2$ 
in Section~\ref{sec:sigmapidwa}. The branching puzzle defined
in Section~\ref{sec:brocedures} is used later in Section~\ref{sec:jhuyhjy}
for the lower bound for monadic~$\Sigma_1$ (Section~\ref{sec:jhuyhjy}).

\subsection{Deterministic tiling}\label{sec:procedures}

Our (deterministic) {\it tiling puzzle\/} is defined as a~quadruple 
$$\G=\<\T,\R,\rE,\ok\>,$$
\noindent where $\T$ is a~finite
a~set of {\it tiles\/}, $\R:\T^4\to\T$ is a {\it tiling function\/},
and $\rE$, $\ok$ are different elements of~$\T$. 
Such $\G$  defines a~unique {\it tiling\/}
\mbox{$\TG:\NN\times\NN\to\T$}, as follows:
\begin{itemize}
\item $\TG(m,n)=\rE$, when $n=0$ or $m=0$;
\item $\TG(m{+}1,n{+}1)=\R({\rm K,L,M,N})$, where\\
${\rm K}=\TG(m,n{+}1)$, ${\rm L}=\TG(m,n)$,
${\rm M}=\TG(m{+}1,n)$,
and ${\rm N}=\TG(m{+}2,n)$. 
\end{itemize}

\begin{figure}[h]
\begin{center}\begin{tikzpicture}[scale=0.7]
\draw (0,0) -- (3,0);\draw (0,1) -- (3,1);
\draw (0,2) -- (1,2);
\foreach \i in {2,...,3} {\draw (\i,0) --(\i,1);};  
\foreach \i in {0,...,1} {\draw (\i,0) --(\i,2);};  
\draw[dashed] (1,2) -- (2,2) -- (2,1);
\node at (.5,.5) {L};\node at (.5,1.5) {K};
\node at (1.5,0.5) {M};\node at (2.5,0.5) {N};
\node at (1.5,1.5) {T};
 \node at (.5,-.5){\scriptsize$m$};\node at (1.5,-.5){\scriptsize$m{+}1$};
\node at (2.5,-.5){\scriptsize$m{+}2$};\node at (-.5,1.5){\scriptsize$n{+}1$};
\node at (-.5,.5){\scriptsize$n$};
\node at (4,0){~};
\end{tikzpicture}
\end{center}
\vspace*{-4ex}
\caption{Result tile.}
\label{fig:tile-result}
\end{figure}
\noindent That is, the  tile $\rE$ is placed along the horizontal and 
vertical edges of the grid $\NN\times\NN$ 
and every other tile is determined by its neighbourhood
consisting of four tiles: one tile to the left and three tiles below. This is 
illustrated by Fig.~\ref{fig:tile-result}, where ${\rm T}=\R({\rm K,L,M,N})$.%
We say that $\G$ is {\it solvable\/} when 
$\TG(m,n)=\ok$, for some numbers $m,n$. The following is unavoidable:

\begin{lem}\label{lemma:un-solvability}
It is undecidable to determine if a given tiling puzzle is solvable.
\end{lem}
\proof  A routine reduction of the following
  problem:\smallskip

\noindent\hfil 
 {\it Does a deterministic Turing Machine accept the empty input?}%
\smallskip

\noindent Row $n$ in the tiling corresponds to the $n$-th step of 
a computation.
\relax\qed

\noindent {\it Locations in tilings:} 
Let $\Pos{(m,n)}=\{(k,l)\ |\ l\leq n\wedge k\leq m+n-l\}$. 
To place a tile at a~location $(m,n)$, where $m,n>0$,
we must tile all locations in $\Pos{(m,n)}$, as 
illustrated in~Figure~\ref{fig:dependency-of-locations},
where the gray square is the location~$(m,n)$.
Define $(m,n)\preceq (k,l)$ when $\Pos(m,n)\subseteq \Pos(k,l)$.

\begin{lem}\label{hfhawqwews} 
The relation $\preceq$ is a well-founded partial order.\qed
\end{lem}

\begin{figure}[h]
\begin{center}
\begin{tikzpicture}[scale=.4]
\foreach \i in {0,...,9} {\draw (\i,0) --(\i,7);};  
\foreach \j in {0,...,7} {\draw (0,\j) -- (9,\j);};
\draw (9,0) -- (15,0);
\draw (9,6) -- (10,6); \draw (9,5) -- (11,5); 
\draw (9,4) -- (12,4); \draw (9,3) -- (13,3); 
\draw (9,2) -- (14,2); \draw (9,1) -- (15,1);
\draw (10,6) -- (10,0); \draw (11,5) -- (11,0); 
\draw(13,3) -- (13,0); \draw(12,4)-- (12,0); 
\draw(14,2)-- (14,0); \draw (15,1)-- (15,0);
\draw [fill=lightgray] (8,6) rectangle (9,7);
\end{tikzpicture}
\end{center}
\vspace*{-3ex}
\caption{Dependency of locations.}
\label{fig:dependency-of-locations}
\end{figure}

\subsection{Branching puzzle}
\label{sec:brocedures}

\noindent
We now generalize our definition of a tiling puzzle to account 
for branching (aka universal) computation,
a~phenomenon dual to nondeterminism. This will be 
needed in Section~\ref{sec:jhuyhjy}. 
A~{\it branching Turing Machine\/}
may divide its computation into multiple branches, each 
proceeding independently.  The whole computation 
can thus be seen as a~tree. We can imagine that every branch follows
its own time line, so we deal with a~tree-like, rather than linear,
flow of computation.

The machine accepts when all these branches
reach accepting states. For simplicity we assume that a branching 
machine divides the computation at {\it every\/} step (there are 
no ordinary deterministic states) and always into two: 
there is always a ``left'' and a ``right'' development.
 Therefore every computation branch can be identified 
by a sequence of binary choices. 
This resembles very much the behaviour of a~deterministic tree automaton: 
the sequence of moves along any fixed branch is fully unique.
In this respect, a branching machine is just a~deterministic  
machine operating in a branching environment.

A {\it branching puzzle\/} is defined again as a tuple of the form 
$$\G=\<\T,\R,\rE,\ok\>,$$ 
with the only difference that 
now the tiling function is $\R:\T^4\to\T^2$. The {\it tiling\/} 
defined by~$\G$ is a~function $\TG:\NN\times\{0,1\}^*\to\T$,
that is, the space to be tiled is $\NN\times\{0,1\}^*$. 
One can imagine a tiling 
of $\NN\times \{0,1\}^*$ as a full binary tree labeled by rows 
of tiles (the label of a node $w\in\{0,1\}^*$ is the sequence 
of tiles $\TG(n,w)$, for all~$n$). This tree represents
a universally branching computation with all possible sequences 
of binary choices. The definition follows:
\begin{itemize}
\item $\TG(n,w)=\rE$, when $n=0$ or $w=\varepsilon$. 
\item $\TG(m{+}1,wi)=\pi_i(\R({\rm K}_i\rm{,L,M,N}))$, 
for $i=0,1$, where\\
${\rm K}_i=\TG(m,wi)$, ${\rm L}=\TG(m,w)$,
${\rm M}=\TG(m{+}1,w)$,
and ${\rm N}=\TG(m{+}2,w)$;
\end{itemize}
A tiling $\TG$ determines, for every infinite path $\pi$ 
in the tree $\{0,1\}^*$, a tiling $\TG_\pi$ of~\mbox{$\NN\times\NN$}, given by 
$\TG_\pi(m,n)=\TG(m,w)$, where $|w|=n$ and $w\subseteq \pi$.
We call it a {\it local tiling\/} \mbox{associated with~$\pi$}. 

Let $s\in\NN$. The puzzle $\G$ is $s$-{\it solvable\/} iff, for every $w$
with $|w|=s$, there is a~prefix $w'$ of~$w$ and a number $m\leq s$ 
such that $\TG(m,w')=\ok$.  
That is, an~$\ok$ tile must be reached at every branch
of the tree of length $s$ and
 it must be at most the $s$-th tile in the row.  

For technical reasons we also need a relativized notion of
$s$-solvability. We say that $\G$ is 
$s$-{\it solvable from\/}~$v$ %
when, for every $w$ such that $v\subseteq w$ and $|w|=s$,
there is a pair $(m,w')$ with $m\leq s$,
 $w'\subseteq w$, and $\TG(m,w')=\ok$. (Then either
$w'\subseteq v$ or $v\subseteq w'\subseteq w$.) 
We have: 

\begin{lem}\label{taksci}
A branching puzzle $\G$ is $s$-solvable iff it is 
$s$-solvable from~$\varepsilon$. It is $s$-solvable from 
a word $v$ with $|v|<s$ \iff it is $s$-solvable from both~$v0$ and~$v1$.\qed
\end{lem}

Here is an analogue of Lemma~\ref{lemma:un-solvability}:

\begin{lem}\label{unavoiis}
The following problem is complete for universal exponential time
(that is, \mbox{co-{\sc Nexptime}-complete}): {\rm Given a branching 
time puzzle $\G$
and a number $s$ (written in binary)  determine 
if $\G$ is $s$-solvable.} 
\end{lem}
\proof  Fix a branching Turing Machine $M$ working in time
$2^{n^k}$ and an input word $a_1a_2\ldots a_n$. Recall that  
all states of the machine are universal and the computation splits into 
two at each step. The encoding of the machine 
is quite natural: the number $s$ is $2^{n^k}$ (this takes~$n^k$ space 
in binary) and the set of tiles is $\T_M = \{\rE,\ok\}\cup\Sigma
\cup (\Sigma\times Q)$, where $\Sigma$ is the machine alphabet and $Q$ is
the set of states. Suppose for example 
that the machine divides the computation making these two moves
when scanning $a$ in state~$q$:

-- write $b$, move left, go to state $p$;

-- write $c$, move right, go to state $r$.

\noindent Fix some $*\in\Sigma$. 
We may now define
$\R(x,x,y,(a,q))\,{=}\,\<(y,p),y\>$, $\R(b,(a,q),y,z)=\<y,*\>$, 
\mbox{$\R((y,p),y,(a,q),z)\,{=}\,\<b,*\>$},
$\R(x,x,(a,q),z)=\<*,c\>$, 
$\R(c,(a,q),y,z)=\<*,(y,r)\>$, and  $\R(\<y,r\>,y,x,z)=\<x,x\>$, 
and $\R(x,x,y,z)=\<y,y\>$, for every $x,y,z\in\Sigma$ (assuming
that $b\neq c$). 
Note that $*$ can be arbitrary --- this value is irrelevant. 
The definition of $\R$ must also ensure the proper positioning 
of tiles representing the input (in row number~$1$), etc. 
By induction \wrt~$(w,n)$ one proves that $\TG(n,w)$ is the content
of tape cell $n$ at time $|w|-1$ in one of the parallel computation 
branches.\qed

\section{Undecidability for $\Delta_2$}
\label{sec:sigmapidwa}

\noindent
We encode a tiling puzzle $\G=\<\T,\R,\rE,\ok\>$ as a~$\Delta_2$~formula
$\Phi_\G$ over the signature:
\begin{itemize}
\item nullary symbols: $\start$, $\lupa$;
\item unary relation symbol $\pula$; 
\item unary relation symbols $\rT$, for each tile $\rT\in\T$;
including $\rE$;
\item unary relation symbols $\rA$, $\rB$, representing border
positions; 
\item binary relation symbols $\rH$, $\rV$, representing horizontal
and vertical neighbourhood.
\end{itemize}
The intuition is that object variables occurring in formulas may 
be interpreted as tile locations. Then $\rA(x)$ can be read as 
``$x$ belongs to the bottom row'' and $\rB(x)$ as  ``$x$ belongs 
to the leftmost column''.\footnote{We use ``Asphalt'' and ``Barrier'' 
as mnemonics.}
The intuitive meaning of $\rH(x,y)$ is 
``$x$ is to the left of $y$'' and $\rV(x,y)$ is understood as 
``$x$ is below $y$''. 

Below we show that $\G$ is solvable \iff~$\Phi_\G$ has a proof. 
The reader has to be aware that the argument to follow
is  proof-theoretical rather than semantical. We are not concerned with the 
interpretation of our formulas in any model, but in their formal structure 
and in the mechanism of proof search. Every step in the 
construction of the tiling is encoded by an expansion of the 
proof environment:~adding new tiles corresponds to adding more assumptions.

Let $\Niedelta$ be a set of formulas in the above signature.
According to the intuition above, predicates $\rH$ and  $\rV$ 
may determine the coordinates $(m,n)$ of a variable $x$ in the grid. 
In general, this is not always consistent, i.e., a~variable~$x$ 
may have one or more pairs of {\it coordinates in\/}~$\Niedelta$.
We define it more formally by induction~\wrt~$(m,n)$. 
\begin{itemize}
\item If $\rA(x),\rB(x)\in\Niedelta$ then $x$ has coordinates $(0,0)$.
\item If $\rH(x,y)\in\Niedelta$ and $x$ has coordinates $(m,n)$
then $y$ has coordinates $(m+1,n)$.
\item If $\rV(x,y)\in\Niedelta$ and $x$ has coordinates $(m,n)$
then $y$ has coordinates $(m,n+1)$.
\end{itemize}
A finite set of formulas (i.e., an environment) $\Niedelta$ is {\it good\/} 
when all formulas in $\Niedelta$ are of the forms $\rA(x)$, $\rB(x)$, 
$\rH(x,y)$, $\rV(x,y)$, or ~$\rT(x)$, where $\rT\in\T,$ and in addition:
\begin{itemize}
\item Each $x\in \FV(\Niedelta)$
has exactly one pair of coordinates.
\item For each $x\in \FV(\Niedelta)$ with coordinates $(m,n)$, 
and every $\rT\in \T$, 
\begin{itemize}
\item $\rT(x)\in\Niedelta$ \iff~$\TG(m,n)=\rT$; 
\item $\rB(x)\in\Niedelta$ \iff~ $m=0$;
\item $\rA(x)\in\Niedelta$, \iff~$n=0$.
\end{itemize}
\end{itemize}
The intuition is that a good environment consistently represents 
partial information about the tiling~$\TG$,  with possible redundancy:
several variables may have the same coordinates.

The formula $\Phi_\G$ to be constructed is of the form 
$\zeta_1\to\cdots\to\zeta_m\to \start$, where some~$\zeta_i$ are 
in $\Pi_1$ and others are in $\Sigma_1$. 
 Technically, it is convenient to define the environment  
\mbox{$\Gamma_\G=\{\zeta_i\ |\ i=1,...,m\}$}
and consider the entailment problem $\Gamma_\G\vdash\start$.
For every ``rule'' of the form \mbox{$\R({\rm K,L,M,N})={\rm T}$}, the set 
$\Gamma_\G$ contains the formula:
\def\labelenumi{{\rm(\arabic{enumi})}}
\begin{enumerate}\setcounter{enumi}{-1}
\item\label{forzero}
 $\quad\forall xyzuv\,
[\rK(y)\,{\to}\,\rL(z)\,{\to}\,\rM(u)\,{\to}\,\rN(v)\,{\to}\,
\rV(z,y)\,{\to}\,\rH(z,u)\,{\to}\,\rH(u,v)\,{\to}\,$\\
\nictu\hfill
 $(\rT(x)\,{\to}\,\rH(y,x)\,{\to}\,\rV(u,x)\,{\to}\,\lupa)\,
{\to}\,\pula(x)].\qquad$
\end{enumerate}

\noindent 
The intended meaning of the formula~($\ref{forzero}$)
is illustrated by Figure~\ref{fig:trzecia}. 
Variables $xyzuv$  represent tile positions, and the 
assumptions $\rK(y),\dots,\rH(u,v)$ describe the situation in the
tiling before placing tile $\rT$ at~$x$. Formula~($\ref{forzero}$)
provides a proof tactic which can be used towards a goal of the 
form $\pula(x)$ as follows.
Find some $yzuv$ witnessing $\rK(y),\dots,\rH(u,v)$, and
prove $\lupa$ under the additional assumptions
$\rT(x),\,\rH(y,x),\,\rV(u,x)$ which  
extend the proof environment to account for the new tile.
\begin{figure}[h]
\begin{center}
\begin{tikzpicture}[scale=0.7]
\draw (0,0) -- (3,0);\draw (0,1) -- (3,1);
\draw (0,2) -- (1,2);
\foreach \i in {2,...,3} {\draw (\i,0) --(\i,1);};  
\foreach \i in {0,...,1} {\draw (\i,0) --(\i,2);};  
\draw[dashed] (1,2) -- (2,2) -- (2,1);
\node at (.5,.5) {L};\node at (.5,1.5) {K};
\node at (1.5,0.5) {M};\node at (2.5,0.5) {N};
\node at (1.5,1.5) {T};
\node at (.5,-.5){$z$};\node at (1.5,-.5){$u$};
\node at (2.5,-.5){$v$};
\node at (.5,2.5){$y$};\node at (1.5,2.5){$x$};
\end{tikzpicture}\hspace{1cm}\nictu
\end{center}
\vspace*{-4ex}
\caption{Formula~$(\ref{forzero})$.}
\label{fig:trzecia}
\end{figure}

\noindent The other formulas in $\Gamma_\G$ are listed below. 
Observe that all quantifiers 
in formulas~\mbox{(\ref{forzero},\ref{fordwa}--\ref{forsze})} are positive, 
while~$\forall x$ in~$(\ref{forjed},\ref{forsie})$ are negative,
and that in the formula~$\Phi_\G$ all signs are reversed.
Since there is no alternation of signs, we obtain that $\Phi_\G$
belongs to $\Delta_2$\,.
\begin{enumerate}
\item\label{forjed} $\quad\forall x\,(\rE(x)\,{\to}\, \rA(x)\,{\to}\, 
\rB(x)\,{\to}\,\lupa)\,{\to}\,\start$;
\item\label{fordwa}
$\quad\forall x\forall y\,(\rE(y) \,{\to}\, \rA(y)\,{\to}\, 
(\rH(y,x)\,{\to}\,\rE(x)\,{\to}\rA(x)\,{\to}\,
\lupa)\,{\to}\,\pula(x))$;
\item\label{fortrzy} 
$\quad\forall x\forall y\,(\rE(y)\,{\to}\,\rB(y) \,{\to}\, 
(\rV(y,x)\,{\to}\,\rE(x)\,{\to}\rB(x)
\,{\to}\,\lupa)\,{\to}\,\pula(x))$;
\item\label{forsze} $\quad\forall x\,(\ok(x)\to\lupa)$;
\item\label{forsie} $\quad\forall x\,\pula(x)\to\lupa$.
\end{enumerate}
The basic machinery here is as follows: to prove $\lupa$ using 
(\ref{forsie}) one needs to derive $\pula(x)$, for a fresh~$x$.
This can be done using one of the proof tactics 
(\ref{forzero},\ref{fordwa},\ref{fortrzy}). Each of these tactics 
verifies some conditions, adds more assumptions, and brings back 
the proof goal $\lupa$. The iteration is started by an attempt to prove 
$\start$ using~(\ref{forjed}). An assumption of the form $\ok(x)$ can be
used to stop the iteration by applying~(\ref{forsze}). 
Before we state the next lemma, let us observe that 
good environments only consist of atoms, and targets of non-atomic formulas 
in~$\Gamma_\G$ are $\start$, $\lupa$, and $\pula$. 
Suppose that $\Niedelta$
is good and that $\alpha$ is a unary or binary atom other than $\pula$.
It follows from 
Lemma~\ref{lemma:lnfgggrd} that $\Gamma_\G,\Niedelta\vdash \alpha$ 
is only possible  when $\alpha$ actually belongs to~$\Niedelta$.

\begin{lem}\label{tami1}
If $\Gamma_\G,\Niedelta\vdash P:\lupa$, for some good~$\Niedelta$, and 
some long normal proof~$P$, then $\G$ is solvable.
\end{lem}
\proof  We proceed by induction \wrt~the length of~$P$. 
  Since $\lupa$ is an atom,
the long normal proof $P$ must begin with a~proof variable~$Y$
declared in $\Gamma_\G,\Niedelta$ so that its type ends with~$\lupa$
(cf.~Lemma~\ref{lemma:lnfgggrd}).
If $Y$ is of type~(\ref{forsze}) then $P=Yx'D$, where~$x'$ is an object variable
and $\Gamma_\G,\Niedelta\vdash D:\ok(x')$. Then $\ok(x')$ must 
actually be in~$\Niedelta$. Hence $\TG(m,n)=\ok$, for some~$m,n$. 

Otherwise $Y$ is of type~(\ref{forsie}) and $P=Y(\lambda x'\,F)$ with 
$\Gamma_\G,\Niedelta\vdash F:\pula(x')$ and $x'$ not free in the environment
$\Gamma_\G,\Niedelta$.
Again, the term $F$ must begin with a variable $Z$ declared 
in $\Gamma_\G,\Niedelta$.
If~$Z$ is of type~(\ref{forzero}) then 
$F= Zx'y'z'u'v'D_{\rK}D_{\rL}D_{\rM}D_{\rN}D_{\rV}D^1_{\rH}D^2_{\rH}(\lambda 
Z_1Z_2Z_3.\,D)$,
where:
\begin{itemize}
\item Terms $D_{\rK}$, $D_{\rL}$, $D_{\rM}$, $D_{\rN}$, $D_{\rV}$, 
$D^1_{\rH}$, $D^2_{\rH}$ are respectively of types 
$\rK(y')$, $\rL(z')$, $\rM(u')$, $\rN(v')$, $\rV(z',y')$, $\rH(z',u')$,
$\rH(u',v')$ in the environment $\Gamma_\G,\Niedelta$; 
\item 
\mbox{$\Gamma_\G,\Niedelta,Z_1\,{:}\,\rT(x'),Z_2\,{:}\,\rH(y',x'),
Z_3\,{:}\,\rV(u',x') \vdash D\,{:}\,\lupa$};
\item $\rT=\R({\rm K,L,M,N})$.
\end{itemize}
But if a long normal form has type $\rK(y')$ in $\Gamma_\G,\Niedelta$
then it must be a proof variable. The same holds for all 
the proofs mentioned in the first item above: these atoms 
must simply belong to~$\Niedelta$. Since $\Niedelta$ is good, we
have $\TG(m+1,n+1)=\rT$. 

Let $\Niedelta'=\Niedelta,\rT(x'),\rH(y',x'),\rV(u',x')$.
The environment~$\Niedelta$ is good, so the variables $y'$,~$z'$, and $u'$
have only one pair of coordinates each. In addition, the presence 
of  assumptions $\rV(z',y')$ and  $\rH(z',u')$ forces that the 
coordinates of  $y'$, $z'$, $u'$ are of the form 
$(m,n+1)$, $(m,n)$, and $(m,n+1)$, respectively. Since 
$\rH(y',x'),\rV(u',x')\in \Niedelta'$, the added variable $x'$ 
has coordinates $(m+1,n+1)$ in~$\Niedelta'$, and this is the only such pair. 
It follows that $\Niedelta'$ is a~good environment, and we can apply 
induction to $D$ because it is a proof of $\lupa$ shorter than $P$. 

Now suppose that $\Gamma_\G,\Niedelta\vdash F:\pula(x')$, 
where the long normal proof~$F$ begins 
with a variable~$Z$ of type~(\ref{fordwa}). Then 
$F=Zx'y'D_ED_A(\lambda Z_1Z_2Z_3.\,D)$, where $D_E$ and $D_A$ are,
respectively, of type $\rE(y')$ and $\rA(y')$, and\smallskip

\noindent\hfil
$\Gamma_\G,\Niedelta,Z_1:\rH(y',x'), Z_2:\rE(x'), Z_3:\rA(x')\vdash D:\lupa$.
\smallskip

\noindent
As in the previous case, the atoms 
$\rE(x')$ and $\rA(x')$ must occur in $\Niedelta$.
To apply induction it suffices to prove that the 
environment\smallskip

\noindent\hfil $\Niedelta'=\Niedelta,Z_1:\rH(y',x'), Z_2:\rE(x'), Z_3:\rA(x')$
\smallskip

\noindent is good. 
Since $\Niedelta$ is good, the variable $y'$ has exactly one pair 
of coordinates~$(m,0)$. The new variable $x'$ has the coordinates 
$(m+1,0)$ and this is its only pair of coordinates. We conclude 
that $\Niedelta'$ is good. 

A long normal proof of $\pula(x')$ may also begin with a variable of 
type~(\ref{fortrzy}). Then the argument is similar as in 
case~(\ref{fordwa}). \qed

\noindent From Lemma~\ref{tami1} we immediately obtain: 

\begin{lem}\label{tami2}
If $\Gamma_\G\vdash\start$ then $\G$ is solvable.
\end{lem}
\proof  A long normal proof of $\start$ must be of the form 
\mbox{$D=Z(\lambda x\lambda XY\ciut V.\,D')$}, for some variable~$Z$ 
of type~(\ref{forjed}) and some $D'$ with\smallskip

\hfil \mbox{$\Gamma_\G,X\,{:}\,\rE(x),Y\,{:}\,\rA(x), V\,{:}\,\rB(x)
\vdash D':\lupa$}.\smallskip

\noindent The set $\Niedelta=\{\rE(x), \rA(x), \rB(x)\}$ 
is good and we apply Lemma~\ref{tami1}.
\relax\qed

Our next aim is to show the converse to Lemma~\ref{tami2}. For the rest
of this section we assume that $\G$ is solvable 
with $\TG(m_0,n_0)=\ok$.
For a good set $\Niedelta$ define
\smallskip

\hfil $S_\Niedelta=\{(m,n)\ |\ \mbox {\rm some } x\in \FV(\Niedelta)\ 
\mbox{\rm has coordinates}\ (m,n)\}$. \smallskip

\noindent
We say that a~set~$\Niedelta$ of formulas is {\it very good\/}
when $\Niedelta$ is good and:
\begin{itemize}
\item The set $S_\Niedelta$ is a subset of $\Pos(m_0,n_0)$;
\item For every $(m,n)\in S_\Niedelta$, exactly 
one $x\in\FV(\Niedelta)$ has coordinates  $(m,n)$;
\item If $x\in\FV(\Niedelta)$ has coordinates  $(m+1,n)$
then some $\rH(y,x)$ is in $\Niedelta$;
\item If $x\in\FV(\Niedelta)$ has coordinates  $(m,n+1)$
then some $\rV(y,x)$ is in $\Niedelta$.
\end{itemize}
A very good set represents the tile assignment without redundancy, 
and every non-zero location is ``justified'' by 
its neighbours occurring below and to the left of it. 

\begin{lem}\label{nazad}
If $\Niedelta\neq\pusty$ is very good then $\Gamma_\G,\Niedelta\vdash\lupa$. 
\end{lem}
\proof  The proof is by induction \wrt~the cardinality of the set 
\mbox{$\Pos(m_0,n_0)-S_\Niedelta$}. In the base case we
have $(m_0,n_0)\in S_\Niedelta$, whence $\ok(x)\in\Niedelta$, for some~$x$.
We use the assumption~(\ref{forsze}) to derive $\lupa$.

For the induction step, 
let $(m',n')\in \Pos(m_0,n_0)-S_\Niedelta$ be minimal \wrt~$\preceq$ 
(cf.~Lemma~\ref{hfhawqwews}). 
Assume first \relax
    that $m'=m+1$ and $n'=0$. By the minimality of~$(m',n')$, 
    there is a unique variable $y\in\FV(\Niedelta)$ with coordinates $(m,0)$
    and with $\rE(y),\rA(y)\in\Niedelta$. Take a fresh variable~$x$. 
    Then $\Niedelta'=\Niedelta\cup\{\rH(y,x),\rE(x),\rA(x)\}$ is very good,
    whence $\Niedelta'\vdash\lupa$. That is, we have 
\mbox{$\Gamma_\G,\Niedelta\vdash 
\rH(y,x)\,{\to}\,\rE(x)\,{\to}\rA(x)\,{\to}\,\lupa$}.
 Using the assumption~(\ref{fordwa}) we 
derive \mbox{$\Gamma_\G,\Niedelta\vdash\pula(x)$}. 
Since $x\not\in\FV(\Niedelta)$, 
we can generalize over~$x$ and obtain 
\mbox{$\Gamma_\G,\Niedelta\vdash\forall x\,\pula(x)$}. 
	Now we use the  assumption (\ref{forsie})
    	to obtain $\Gamma_\G,\Niedelta\vdash\lupa$.

The case $m'=0$ and $n'=n+1$ is similar but  we use assumption~(\ref{fortrzy}).
Assume therefore that
$m'=m+1$ and $n'=n+1$, for some $m,n$. By the minimality
of~$(m',n')$, there are variables $y,z,u,v\in\FV(\Niedelta)$ with coordinates
$(m,n+1)$, $(m,n)$, $(m+1,n)$, $(m+2,n)$. These variables are unique because 
$\Niedelta$ is very good. Also we have 
\mbox{$\rK(y),\rL(z),\rM(u),\rN(v)\in\Niedelta$}, for some unique choice of 
$\rK,\rL,\rM,\rN$. In addition, since $\Niedelta$ is very good, 
we must also have 
in~$\Niedelta$ the assumptions $\rV(z,y),\rH(z,u),\rH(u,v)$.
Let $\Niedelta'=\Niedelta\cup\{\rT(x), \rH(y,x), \rV(u,x)\}$, where~$x$ is 
a~fresh variable, and  
$\rT=\R(\rK,\rL,\rM,\rN)$. Then $\TG(m',n')=\rT$. 
The environment~$\Niedelta'$ is very good, because $x$
is the unique variable with 
coordinates $(m+1,n+1)$. By the induction hypothesis, 
$\Gamma_\G,\Niedelta'\vdash\lupa$, whence 
\mbox{$\Gamma_\G,\Niedelta \vdash 
\rT(x)\,{\to}\,\rH(y,x)\,{\to}\,\rV(u,x)\,{\to}\,\lupa$}.
Using the assumption~(\ref{forzero}) we can now 
derive $\Gamma_\G,\Niedelta\vdash\pula(x)$. 
But we actually have  $\Gamma_\G,\Niedelta \vdash \forall x\,\pula(x)$,
because $x$ is not free in $\Gamma_\G,\Niedelta$. Hence 
$\Gamma_\G,\Niedelta\vdash\lupa$ by an
application of~(\ref{forsie}). 
\qed

\begin{lem}\label{nazad2}
If $\G$ is solvable then $\Gamma_\G\vdash\start$.
\end{lem}
\proof  The set $\Niedelta=\{\rE(x),\rA(x),\rB(x)\}$ 
is very good, so \mbox{$\Gamma_\G,\rE(x),\rA(x), \rB(x)\vdash\lupa$}
holds by Lemma~\ref{nazad}. Hence 
$\Gamma_\G\vdash\rE(x)\to\rA(x)\to\rB(x)\to\lupa$. Using~(\ref{forjed}) 
one derives $\Gamma_\G\vdash\start$.
\relax\qed

\begin{thm}\label{nazaditam} Provability in $\Delta_2$ is undecidable.
\end{thm}
\proof  By Lemma~\ref{lemma:un-solvability}, solvability of tiling 
puzzles is undecidable. Lemmas~\ref{tami2} and~\ref{nazad2} 
give an effective
reduction from the tiling puzzle problem to provability. 
\qed

\subsection*{A finite signature}
Observe that our proof of Theorem~\ref{nazaditam} uses as many predicate
symbols as there are tiles, i.e.,~it applies to an infinite signature.
We now briefly explain how it can be adjusted to work for a finite 
language. First, redefine the tiling puzzle so that 
\mbox{$\G=\<\T,\R,\rE,n,\rT_1,\ldots,\rT_n,\ok\>$}, where $n\in \NN$ and 
$\rT_1,\ldots,\rT_n\in \T$ (possibly with repetitions). This is 
to account for a non-empty input word.  
Require that the tiling satisfies $\TG(i,0)=\rT_i$,
for all $i=1,\ldots,n$. Using a universal Turing Machine (which has
a~fixed number of states and uses a fixed alphabet) prove that for some
$M\in\NN$ the problem of solvability is undecidable for the modified 
puzzles with at most~$M$ tiles. This reduces the number of necessary
predicates to a finite level. The remaining construction is essentially the
same, but one has to adjust formula~(\ref{forjed}) as follows:\label{tufinsyg}
\begin{itemize}   
\item 
$\forall x_0\ldots x_n\,(\rE(x_0)\,{\to}\,  
\rB(x_0)\,{\to}\,\rA(x_0)\,{\to}\cdots{\to}\,\rA(x_n)
\,{\to}\,\rT_1(x_1)\,{\to}\cdots{\to}\,\rT_n(x_n)
\,{\to}$\\
\nictu\hfill$\,\rH(x_0,x_1)\,{\to}\cdots{\to}\,\rH(x_{n-1},x_n)\,{\to}\,
\lupa)\,{\to}\,\start$.
\end{itemize}

\subsection{Monadic $\Delta_2$}\label{subsec:mon}

Our proof of Theorem~\ref{nazaditam} used 
binary relation symbols. We now show how to eliminate them by
a syntactic translation. This is possible, because we only used 
formulas of a very simple shape. 
We say that a formula $\varphi$ is {\it easy\/} when it is 
an atom, or when ${\it target\/}(\varphi)$ is unary or nullary and 
one of the following holds:
\begin{itemize}
\item $\varphi=\forall x\,\psi$, where $\psi$ is easy;
\item $\varphi=\psi\to\vartheta$, where $\psi$ and $\vartheta$  are easy.
\end{itemize}
Observe that the set $\Gamma_\G$ in Section~\ref{sec:sigmapidwa} 
consists of easy formulas.

Let $\rj$ and $\rd$ be fresh 
unary relation symbols (i.e., not occurring in the source language).
With every binary relation symbol~$\rP$ we associate another  fresh
nullary symbol~$\rp$.
We define $\overline{\rP(x,y)} = \rj(x)\to\rd(y)\to \rp$, for 
binary $\rP$, and $\overline{\rP(x)}=\rP(x)$, 
$\overline{\rP}=\rP$, when $\rP$ is unary or nullary. 
Then, by induction, define $\overline{\forall x\,\varphi}=
\forall x\,\overline\varphi$, and $\overline{\varphi\to\psi}=
\overline\varphi\to\overline{\psi}$. 

\begin{lem} The translation $\psi \mapsto \overline{\psi}$ has the 
following properties: 
\begin{itemize}
\item $\mbox\FV(\overline\psi)=\mbox\FV(\psi)$;
\item $\overline{\psi[y/x]}= \overline{\psi}[y/x]$;
\item If $\overline\varphi=\overline\psi$ then $\varphi=\psi$;
\item If $\psi$ is easy then so is $\overline\psi$.
\end{itemize}
\end{lem}
\proof  Routine induction.
\qed

\begin{lem}\label{ppodyrt}
Let $\Sigma$ consist of binary atoms and let 
targets of all formulas in $\Gamma$ be nullary or unary. 
Then $\overline\Gamma,\overline\Sigma \vdash \overline{\rP(x,y)}$
implies $\rP(x,y)\in\Sigma$. 
\end{lem}
\proof 
We have $\overline\Gamma,\overline\Sigma, \rj(x),\rd(y)\vdash \rp$.
No formula in  $\overline\Gamma$ may end with~$\rp$, thus a~long normal 
proof of~$\rp$ must begin with an element of $\overline\Sigma$:
a~variable of type $\overline{\rQ(u,v)}=\rj(u)\to\rd(v)\to \rqq$. 
Then $\rqq=\rp$, i.e., $\rP=\rQ$, and 
we have $\overline\Gamma,\overline\Sigma,\rj(x),\rd(y)\vdash \rj(u)$
and \mbox{$\overline\Gamma,\overline\Sigma,\rj(x),\rd(y)\vdash \rd(v)$}.
There is no other way to prove $\rj(u)$ but to use the assumption~$\rj(x)$.
Hence, $x=u$, and similarly we also obtain $y=v$. 
Thus, $\rP(x,y)=\rQ(u,v)\in\Sigma$. 
\qed

\begin{lem}\label{lem:from-overline-to-normal}
If  $\,\overline\Gamma\vdash\overline\varphi$, where $\varphi$ and 
all formulas in $\Gamma$ are easy, then  $\Gamma\vdash\varphi$.
\end{lem}
\proof 
A {\it quasi-long eliminator\/} is a term of the form $XE_1\ldots E_m$,
where $X$ is a~proof variable and every $E_i$ is either an lnf or an
object variable. Observe that if $\overline\Gamma\vdash M:\tau$, where
$M$ is a~quasi-long eliminator, then either $\tau=\overline\varphi$,
for some~$\varphi$, or $\tau=\rd(y)\to \rp$, or $\tau=\rp$, for some
$\rp$ and $y$. In the last two cases, we have 
 $M=M'N_1$ or $M=M'N_1N_2$, with $M':\overline{\rP(x,y)}$, and 
$N_1:\rj(x)$, and~$N_2:\rd(y)$, for some~$x$ and~$y$.

Let now $\overline\Gamma\vdash M:\overline\varphi$, where $M$ is an lnf
or a quasi-long eliminator. We prove
that $\Gamma\vdash\varphi$, by induction \wrt~$M$. The case of a variable
is obvious.

Let $M=\lambda Z.\,N$. Without loss of generality 
we can assume that $\varphi=\psi\to\vartheta$, because 
the case of $\varphi=\rP(x,y)$ follows from Lemma~\ref{ppodyrt}. Then 
$\overline\Gamma, Z\,{:}\,\overline\psi\vdash N:\overline\vartheta$. 
By the induction
hypothesis for $N$ we have $\Gamma,\psi\vdash\vartheta$,
whence $\Gamma\vdash\varphi$.

If $M=\lambda y.\,N$ (where we can assume $y$ is fresh)
then $\overline\varphi=\forall y\,\tau$, which means 
that $\varphi=\forall y\, \psi$ with $\overline\psi=\tau$. We have
$\overline\Gamma\vdash N:\overline\psi$, so $\Gamma\vdash\psi$ and
thus $\Gamma\vdash\varphi$ by generalization.

If $\overline\Gamma\vdash X\vec EN:\overline\varphi$ then 
the type of $X\vec E$ must be of the form $\overline\psi\to\overline\varphi$,
because $\overline\varphi$ is neither of the form $\rd(y)\to\rp$ nor~$\rp$.
By the induction hypothesis, both $\psi\to\varphi$ and $\psi$ are provable,
and so must be~$\varphi$. 

If $\overline\Gamma\vdash X\vec Ey:\overline\varphi$, 
where $y$ is an object variable, then 
$\overline\Gamma\vdash X\vec E:\forall x\,\tau$, for some~$\tau$
with 
$\overline\varphi=\tau[y/x]$. Since $X\vec E$ is a quasi-long
eliminator, we must have 
$\forall x\,\tau=\overline{\forall x\,\psi}=\forall x\,\overline\psi$, and 
$\overline\varphi=\overline\psi[y/x]=\overline{\psi[y/x]}$. Hence 
$\varphi=\psi[y/x]$.
We apply induction to~$X\vec E$.
\qed

The converse to Lemma~\ref{lem:from-overline-to-normal} is obvious. 
Since all formulas 
used in our coding are easy, we can restate Lemmas~\ref{tami2} 
and~\ref{nazad2} using $\overline\Gamma_\G$ instead of $\Gamma_\G$. 
We conclude with:

\begin{thm}\label{jjg6hgytt}
It is undecidable whether a $\Delta_2$ formula
with unary predicates is provable.\qed
\end{thm}

\subsection*{Generalization:} The translation $\varphi \mapsto \overline\varphi$
can be easily generalized to predicates of any fixed arity~$n\geq 2$, 
by introducing $n$ auxilary symbols ${\bf 1}, {\bf 2},\ldots, {\bf n}$ 
and setting 

$$\overline{\rP(x_1,x_2,\ldots,x_n)}=
{\bf 1}(x_1)\to{\bf 2}(x_2)\to\cdots\to{\bf n}(x_n)\to\rp.$$ 

\noindent It is convenient to assume without loss of generality 
that all many-argument predicates are of the same arity $n$.
Then the proof of the following is virtually the same as in the 
binary case.

\begin{proposition}\label{lem:overline-to-normal}
Let $\varphi$ and all formulas in $\Gamma$ be easy.
Then $\Gamma\vdash\varphi$ iff 
$\,\overline\Gamma\vdash\overline\varphi$.
\end{proposition}

\section{{\sc Expspace}-completeness for $\Sigma_1$}
\label{sec:sigmajeden}

The lower bound is obtained by encoding the halting problem for  bus
machines~\cite{urzy09} 
into the entailment problem for~$\Sigma_1$. A~bus machine is an alternating
computing device operating on a~finite word (bus) of~a~fixed length.
At every step the whole content of~the bus is updated according to
one of~the instructions of~the machine. In addition new instructions
may be created each time and those can be used in later steps.
A~precise definition is as follows. 

A~{\it simple switch\/} over a~finite alphabet~$\A$ is
a~pair of~elements of~$\A$, written $a\tam b$. A~{\it labeled switch}
is a~quadruple, written $a\tam b(c\tam d)$, where the simple switch 
$c\tam d$ is the {\it label}. Finally, a~{\it branching switch\/} 
is a~triple, written $a\tam b\times c$.

A~{\it bus machine\/} is a~tuple $\M=\<\A,m,w_0,w_1,\I\>$,
 where~$\A$ is a~finite alphabet, $m>0$ is the {\it bus length\/} of~$\M$
(the length of~the words processed), $w_0$ and $w_1$ are words of~length~$m$ 
over $\A$, called the {\it initial\/} and {\it final word\/}, respectively,
and~$\I$ is a set of~{\it global instructions\/}.

Every global instruction  is an $m$-tuple
$\II=\<I_1,\ldots,I_m\>$ of~sets of~switches. Switches in~$I_i$ are
meant to act on the $i$-th symbol of~the bus. 
It is required that all switches in a~given instruction $\II$ 
are of~the same kind: either all are simple, or all are labeled,
or all are branching. Therefore we classify instructions as 
simple, labeled, and branching. A~{\it local instruction\/} is 
a~special case of a simple instruction with singleton 
sets at all coordinates.

A~{\it configuration\/} of~$\M$ is a~pair $\<w,\J\>$, where 
$w$ is a~word over~$\A$ of~length~$m$, and $\J$ is a~set of~local instructions.
The {\it initial\/} configuration is $\<w_0,\pusty\>$,
and any configuration of~the form $\<w_1,\J\>$ is called {\it final\/}.

Suppose that $\II = \<I_1,\ldots,I_m\>$, and let $w=a_1\ldots a_m$ and
$w'=b_1\ldots b_m$, $w''=c_1\ldots c_m$. 
Transitions of $\M$ according to~$\II$ are defined as follows:
\begin{itemize}
\item If $\II$ is a~simple instruction, and 
for every $i\leq m$ the switch $a_i\tam b_i$ belongs to~$I_i$,
then $\<w,\J\>\To_\M^{\II}\<w',\J\>$;
\item If $\II$ is a labeled instruction and $a_i\tam b_i(c_i\tam d_i)$ belongs 
to~$I_i$, for every 
$i\leq m$,  then $\<w,\J\>\To_\M^{\II}\<w',\J'\>$, where
$\J'=\J\cup\{\<\{c_1\tam d_1\},\ldots,\{c_m\tam d_m\}\>\}$;
\item If $\II$ is a branching instruction, 
and the switch $a_i\tam b_i\times c_i$ is in $I_i$,
for~every~$i\leq m$, then $\<w,\J\>\To_M^{\II}(\<w',\J\>, 
\<w'',\J\>)$.   
(Now the relation $\To_M^{\II}$ has three arguments.)
	     \end{itemize}
The notion of an accepting configuration of a bus machine is 
defined recursively. We say that  a configuration 
$\<w,\J\>$ is {\it eventually  accepting\/} if
it is either a~final configuration, or 
\begin{itemize}
\item  There is a~non-branching instruction~$\II\in\I\,\cup\,\J$, with
\mbox{$\<w,\J\>\To_\M^\II\<w',\J'\>$}, where $\<w',\J'\>$ is eventually 
accepting, or 
\item  There is a~branching instruction~$\II\in\I\cup\J$
such that  $\<w,\J\>\To_\M^\II(\<w',\J\>, \<w'',\J\>)$, where
both $\<w',\J\>$ and $\<w'',\J\>$ are eventually accepting. 
\end{itemize}
The machine $\M$ {\it accepts\/} 
iff the initial configuration is eventually accepting.
As usual with alternating machines, an accepting computation of
a bus machine should be imagined as a~tree with final configurations
at all leaves and branching transitions at branching nodes.%

\begin{example}\label{hhsdsrehdbfgf}\rm
This example, inspired by~\cite{Kusmierek07}, is from~\cite{RehofU12}. 
Let $\A=\{a,b,c,d\}$, and let\smallskip

\hfil $I^+=\{a\,{\tam}\, b(c\,{\tam}\,d)\}$,~~ 
$I^-=\{b\tam a(d\,{\tam}\,c)\}$,\\
\nictu\hfil
$I=\{a\,{\tam}\, a(c\,{\tam}\,c), b\,{\tam}\, b(d\,{\tam}\,d)\}$,
~~$I^*=\{b\,{\tam}\, c\}$\smallskip

\noindent Consider  $\M=\<\A,4,aaaa,dddd,\I\>$, where 
$\I$ consists of~the following tuples:\smallskip

\noindent\hfil$\<I,I,I,I^+\>$, $\<I,I,I^+,I^-\>$, $\<I,I^+,I^-,I^-\>$, 
$\<I^+,I^-,I^-,I^-\>$, $\<I^*,I^*,I^*,I^*\>$.\smallskip

\noindent
The machine $\M$ behaves in a deterministic way, for example the only
instruction applicable in the initial 
configuration $\<aaaa,\pusty\>$
is $\<I,I,I,I^+\>$. Executing it yields $\<aaab,\{I_0\}\>$, where~$I_0$
is the local instruction 
$\<\{c\tam c\},\{c\tam c\},\{c\tam c\},\{c\tam d\}\>$.
The latter can be used later to change a configuration of the
form $\<cccc,\J\>$ into $\<cccd,\J\>$. 
But now the machine must execute $\<I,I,I^+,I^-\>$ and enter 
$\<aaba, \{I_0,I_1\}\>$, where
$I_1=\<\{c\tam c\},\{c\tam c\},\{c\tam d\},\{d\tam c\}\>$.

In the first phase of
computation only global instructions are executed and all words over~$\{a,b\}$
appear on the bus in the lexicographic order. Every application of~a~global 
instruction creates a~new unique local instruction. After arriving at $bbbb$,
the machine rewrites the bus to $cccc$ using $\<I^*,I^*,I^*,I^*\>$
and then executes one by one all 
the local instructions, eventually reaching the final $dddd$. The total number
of steps is $2\cdot 2^4-1$; also 
the number of~local instructions is exponential and so is the (implicit) space 
needed to store them. 

\end{example}

\begin{thm}[{\cite{urzy09}}]\label{ggdsrettry} 
The halting problem for bus machines  (``Does a given machine accept?'')
 is~{\sc Expspace}-complete.\qed
\end{thm}

\noindent
Given a bus machine $\M=\<\A,m,w_0,w_1,\I\>$, we construct (in {\sc Logspace})
a set of universal formulas~$\Gamma_\M$ and an open formula~$\alpha_\M$ 
such that $\Gamma_\M\vdash \alpha_\M$ \iff~$\M$ halts. The free 
variables in~$\Gamma_\M$ and $\alpha_\M$ are identified with the
symbols in~$\A$ and the number, as well as arity, of relation symbols in 
our formulas also depend on~$\M$. The main relation symbol $\bus$
is $m$-ary and it is intended to represent the content of the bus.
The obvious convention is to write $\bus(w)$ for $\bus(a_1,\ldots, a_m)$,
when $w=a_1\ldots a_m$ and $\vec a$ for $a_1a_2\ldots a_m$.

The formula $\alpha_M$ is $\bus(w_0)$, and $\bus(w_1)$ is a member of
$\Gamma_\M$. The idea is that a proof of $\bus(w_0)$ succeeds when 
every branch of a computation can terminate by calling the axiom $\bus(w_1)$.

We associate binary (resp.~ternary, quaternary)
predicate symbols $I$ with sets~$I$ of simple (resp.~branching, labeled)
switches occurring in the instructions of $\M$. Then for every simple
switch $a\tam b$ in $I$, the atomic formula $I(a,b)$ is
placed in $\Gamma_\M$, and similarly for branching and labeled 
switches.
For example, 
the set $I$ in Example~\ref{hhsdsrehdbfgf} yields two assumptions 
$I(a,a,c,c)$ and $I(b,b,d,d)$.

In $\Gamma_\M$ there are also formulas $\psi_\II$ 
for all global 
instructions~$\II$ in~$\I$. 
In~case of a simple instruction $\II=\<I_1,\ldots,I_m\>$,
the formula takes the form:
\begin{enumerate}
\item $\psi_\II=\forall\vec x\vec y\,(I_1(x_1,y_1)\to\cdots\to
I_m(x_m,y_m)\to \bus(\vec y\,)\to \bus(\vec x\ciut))$.\smallskip

If $\II=\<I_1,\ldots,I_m\>$ is a labeled instruction,  then:\smallskip

\item
\begin{tabular}[t]{@{}l@{\,}l@{}}
$\psi_\II=
\forall\vec x\vec y\vec z\vec u\,(I_1(x_1,y_1,z_1,u_1)$
& $\to\cdots\to I_m(x_m,y_m,z_m,u_m)$\\
& $\to((\bus(\vec u\ciut)\to\bus(\vec z\,))\to 
\bus(\vec y\,))\to \bus(\vec x\ciut))$.
\end{tabular}\\\smallskip
\end{enumerate}

\noindent 
Finally, for a branching instruction $\II=\<I_1,\ldots,I_m\>$,
we take:\smallskip

\begin{enumerate}[start=3]
\item
\begin{tabular}[t]{@{}l@{\,}l@{}}
$\psi_\II=
\forall\vec x\vec y\vec z\,(I_1(x_1,y_1,z_1)$
& ${\to\cdots\to} I_m(x_m,y_m,z_m)$ 
 $\to\bus(\vec z\ciut)\to  \bus(\vec y\,)\to\bus(\vec x\ciut)).$
\end{tabular}\smallskip
\end{enumerate}

\noindent
A local instruction $J$ may be identified 
with a~rewrite rule of the form $w\To v$. Such a~rule will be represented
as a formula~$\varphi_J$ of the form $\bus(v)\to \bus(w)$. We define
$\Gamma_\J=\{\varphi_J\ |\ J\in \J\}$. 

To see the motivation, suppose we want to derive 
$\Gamma_\M\vdash \bus(bbbb)$, where $\M$ is as in Example~\ref{hhsdsrehdbfgf}.
We use the formula $\psi_{\<I^*,I^*,I^*,I^*\>}$: 
$$\forall\vec x\vec y(I^*(x_1,y_1)\to I^*(x_2,y_2)\to I^*(x_3,y_3)\to 
I^*(x_4,y_4)\to \bus(\vec y\,)\to \bus(\vec x\ciut)),$$

\noindent instantiated 
by substituting $b$ for $x_i$ and $c$ for $y_i$. Since
the assumption $I^*(b,c)$ is in $\Gamma_\M$, the task of proving
$\bus(bbbb)$ is reduced to proving $\bus(cccc)$.

\begin{lem}\label{masijeden}
A configuration $\<w,\J\>$ is eventually  accepting 
iff~the judgment 
$$\Gamma_\M,\Gamma_\J\vdash\bus(w)$$
is derivable.
\end{lem}
\proof  
From left to right the proof is by induction \wrt~the 
definition of an eventually  accepting configuration. 
Let $\<w,\J\>$ be eventually  accepting. If it is final,
the proof is trivial, because $\bus(w_1)\in\Gamma_\M$.  
Otherwise, assume for example
that \mbox{$\<w,\J\>\To_\M^{\II}\<w',\J'\>$}, where 
\mbox{$\II=\<I_1,\ldots, I_m\>$} is a~labeled instruction, 
and $\<w',\J'\>$ is eventually  accepting.  
Then $\J'=\J\cup\{J\}$, where
$J$ is a new local instruction.
By the induction hypothesis 
we have \mbox{$\Gamma_\M,\Gamma_{\J},\varphi_J\vdash\bus(w')$}.
It follows that 
\mbox{$\Gamma_\M,\Gamma_{\J}\vdash \varphi_J\to\bus(w')$}.
For $j=1,\ldots,m$, let 
\mbox{$a_j\tam b_j (c_j\tam d_j)$} be the switches 
used in this step. Then
$w=a_1\dots a_m$, $w'=b_1\dots b_m$, 
and $\varphi_J=\bus(d_1\dots d_m)\to \bus(c_1\dots c_m)$. 
Hence \mbox{$\Gamma_\M,\Gamma_{\J}\vdash (\bus(\vec d\,)\to 
\bus(\vec c\ciut))\to\bus(\vec b\,)$}. 
We have all the $I_j(a_j,b_j,c_j,d_j)$ in $\Gamma_\M$,  
so we prove $\bus(\vec a)$ using the appropriate axiom~(2)  
instantiated with $\vec x := \vec a$, $\vec y := \vec b$, $\vec z := \vec c$, 
$\vec u := \vec d$. Other cases are similar. 

The proof in the direction from right to left is by induction
\wrt~the length of long normal proofs. Assume that 
$\Gamma_\M,\Gamma_\J\vdash\bus(w)$. If $w$ is not final then a long
normal proof must begin with a variable of type (1),~(2), or (3).
Suppose for example that (3) is the case. For some instantiation
 \mbox{$\vec x := \vec a = w$, $\vec y := \vec b$}, $\vec z := \vec c$, there
are proofs of~$I_i(a_i,b_i,c_i)$ and of $\bus(\vec b\,)$ and 
$\bus(\vec c\ciut)$.
A~proof of~$I_i(a_i,b_i,c_i)$ is only possible
when $I_i(a_i,b_i,c_i)$ actually occurs in $\Gamma_\M$.
This is because there are no other assumptions 
with target~$I_i$. In particular this
proves that variables $b_i$, $c_i$ do correspond to actual bus symbols.
Since $\bus(\vec b)$ and $\bus(\vec c)$ are provable, it follows from the
induction hypothesis that $\<\vec b,\J\>$ and $\<\vec c,\J\>$ are 
eventually  accepting. 
Therefore also $\<w,\J\>$ is eventually  accepting. 
\qed

\subsection[kk]{An upper bound for $\Sigma_1$}

A judgment of the form 
\mbox{$\Gamma\vdash\varphi$}, where $\varphi$ is a~$\Sigma_1$ formula 
and all assumptions
in~$\Gamma$ are $\Pi_1$ formulas, is called 
a {\it $\Sigma_1$ judgment\/}. Observe that normal proofs
of $\Sigma_1$ judgments are of the forms:
\begin{enumerate}[label=\alph*)]
\item $\Gamma\vdash \lambda X\ciut{:}\,\alpha.\,M:\alpha\to\beta$;
\item $\Gamma\vdash XM_1\ldots M_r:\beta$, 
\end{enumerate}
where $M$ is a normal proof term and 
each $M_i$, for $i=1,\ldots,r$, is a~normal proof term or an object variable.
Proofs of shape (b) are called {\it eliminators\/}. 
We say that $N'$ is an {\it instance\/} of~$N$ when
\mbox{$N'=N[\vec x := \vec y\ciut]$}, for some object variables $\vec x,\vec y$.
The following is an easy consequence of Lemma~\ref{lkiojiul}.

\begin{lem}\label{hhfgrtrfg}
Fix an object variable $x_0$, and let 
$\W=\FV(\Gamma)\cup\FV(\varphi)\cup\{x_0\}$.
If \mbox{$\Gamma\,\vdash N:\varphi$} then 
$\Gamma\vdash N':\varphi$, for some instance~$N'$ of~$N$
such that $\FV(N')\subseteq \W$. %
\end{lem}

\proof  Let $\vec x$ be the list of all variables in $\FV(N)-\W$,
and let $\vec y\,$ be any variables in $\W$. (The latter is nonempty 
because of~$x_0$.)  
Then $\Gamma[\vec x := \vec y\ciut]\vdash 
N[\vec x := \vec y\ciut]:\varphi[\vec x := \vec y\ciut]$,
by Lemma~\ref{lkiojiul}. But variables $\vec x$ are neither free in $\Gamma$ 
nor in $\varphi$, whence $\Gamma[\vec x := \vec y\ciut]=\Gamma$ and
$\varphi[\vec x := \vec y\ciut]=\varphi$. 
\relax\qed

Note that if $\FV(\Gamma)\cup\FV(\varphi)\neq\pusty$ then Lemma~\ref{hhfgrtrfg}
yields  
 $\FV(N')\subseteq \FV(\Gamma)\cup\FV(\varphi)$.

\begin{lem}\label{dfdgerdf}
Let \mbox{$\Gamma\,\vdash N:\varphi$}, where~$\Gamma$ 
consists of $\Pi_1$ formulas 
and $N$ is normal. Assume in addition that 
either $N$ is an eliminator or $\varphi$ is a $\Sigma_1$ formula.
Then the term $N$ contains no occurrences of object abstraction.
In addition, if~$N$ is an eliminator then $\varphi$ is in $\Pi_1$.
\end{lem}

\proof  Induction \wrt~$N$. If $N=X$ then the type of $X$ is in 
$\Pi_1$, because $X$ is declared in~$\Gamma$. 

If $N=\lambda X{:}\,\psi.\, P$ then $\psi$ is in $\Pi_1$ 
and \mbox{$\Gamma, X{:}\,\psi\vdash P:\vartheta$}, for some 
$\vartheta\in\Sigma_1$. We use the induction hypothesis for~$P$. 
Case $N=\lambda x\,N'$ is impossible. 
 If $N=X\vec N M$, where $M$ is a~proof term, then we have 
\mbox{$\Gamma\vdash X\vec N:\psi\to\varphi$} and $\Gamma\vdash M:\psi$,
for some~$\psi$. Since $X\vec N$ is an eliminator, the 
formula $\psi\to\varphi$ is in $\Pi_1$ and so must be~$\varphi$,
while~$\psi$ is in $\Sigma_1$. We apply induction to $X\vec N$ and $M$.

Finally, if $N=X\vec N y$, where $y$ is an object variable, then we 
apply induction to~$X\vec N$. 
\qed

\begin{lem}\label{skadzmiennewformule}
If $\Gamma\vdash M:\varphi$ then 
 $\FV(\varphi)\subseteq \FV(\Gamma)\cup\FV(M)$.
\end{lem}
\proof  
Easy induction \wrt~$M$.
\qed

Let $\W$ be a set of variables. If 
$\FV(\Gamma)\cup\FV(\varphi)\cup\FV(M)\subseteq \W$
then we say that $\Gamma\vdash M:\varphi$ 
is a {\it $\W$-judgment}. A judgment is 
{\it $\W$-derivable} when it is derivable using the 
rules in Figure~\ref{prufassirules}
  restricted to $\W$-judgments.

\begin{lem}\label{skadzmiennewdowodzie}
Let $\Gamma\vdash M:\varphi$ be a provable $\W$-judgment.
If $M$ contains no object abstraction then 
$\Gamma\vdash M:\varphi$ is \mbox{$\W$-derivable}.
\end{lem}
\proof  Easy induction \wrt~$M$. In case of application 
one uses Lemma~\ref{skadzmiennewformule}. 
\qed

\begin{lem}\label{jjghty}
The decision problem for $\Sigma_1$ formulas is solvable in {\sc Expspace\/}.
\end{lem}
\proof  To find a proof of a given $\Sigma_1$ formula $\varphi$ 
one uses an obvious generalization of the Ben-Yelles algorithm~\cite{sorm06}
for simple types.
It follows from Lemma~\ref{dfdgerdf} that a normal inhabitant~$N$ of 
a $\Sigma_1$ formula $\varphi$ must not contain any object abstraction.
In addition, by~Lemma~\ref{hhfgrtrfg}, one can assume that free variables of~$N$
are all in the set \mbox{$\W=\FV(\varphi)\cup\{x_0\}$}. 
(The variable~$x_0$ is added to make sure that the set is not empty.)
By Lemma~\ref{skadzmiennewdowodzie}, the judgment $\vdash N:\varphi$ 
is $\W$-derivable. Therefore the algorithm needs only to consider 
judgments $\Gamma'\vdash M:\psi$ where all object variables are 
in~$\W$.  The number of different formulas
in $\Gamma'$ is thus at most exponential in the size $n$ of $\varphi$.
(With at most $n$ variables, every subformula of $\varphi$ has 
at most $n^n$ instances.) Using the same argument as for simple 
types we therefore obtain an alternating exponential time algorithm.
\qed

\begin{thm}\label{hhfawesdess}
The decision problem for $\Sigma_1$ is {\sc Expspace\/}-complete.
\end{thm}
\proof  Lemma~\ref{masijeden} reduces the halting problem
for bus machines to provability in $\Sigma_1$. The upper bound is
provided by Lemma~\ref{jjghty}.\relax
\qed

\section{Arity-bounded~$\Sigma_1$} 
\label{sec:jhuyhjy}

The undecidability of $\Delta_2$ holds even if we require that
all predicates in formulas are unary. Technically, it is the case
because the formulas used in the proof are easy, and we can apply
the translation defined in Section~\ref{subsec:mon}. But the proof
of the {\sc Expspace\/}-hardness of $\Sigma_1$ (Lemma~\ref{masijeden})
uses non-easy formulas of unbounded arity.  

It turns out that the $\Sigma_1$ decision problem is actually 
``easier'' if we set any fixed bound on the arity of formulas. 
For every such bound, in particular in the monadic case, the 
problem turns out only {\it co}-{\sc Nexptime}-complete.

\subsection{The lower bound}\label{ppjytuyt}

To obtain the {\it co}-{\sc Nexptime} lower bound we encode a given branching
puzzle~$\G$ and a constant~$s$ 
as an entailment problem $\Gamma_\G\vdash\start$.  
This is partly similar to the construction in Section~\ref{sec:sigmapidwa},
in particular all formulas in~$\Gamma_\G$ are easy. In addition, all
these formulas are either quantifier-free or universal.
From now on we assume that $s$ is fixed. 
The idea of the encoding can easily be explained if we assume for a while
that the language of arithmetic is in our disposal. Then~$\Gamma_\G$
could be composed of the following assumptions:\medskip

\noindent Tiling step $\G(\rK,\rL,\rM,\rN)=\<\rT,\rU\>$:
\begin{itemize}[label=[$(0_r)$]
\item[$(0_l)$]
 $\forall mt(\rK(m,t{+}1)\,{\to}\,\rL(m,t)\,{\to}\,
\rM(m+1,t)\,{\to}\,\rN(m{+}2,t)$\\
\nictu\hspace{4.5cm}
$\,{\to}\,\lewy(t{+}1)\,{\to}\,(\rT(m{+}1,t{+}1)
\,{\to}\,\lupa)\,{\to}\,\lupa)$.

\item[$(0_r)$]$\forall mt(\rK(m,t{+}1)\,{\to}\,\rL(m,t)\,{\to}\,
\rM(m+1,t)\,{\to}\,\rN(m{+}2,t)$\\
\nictu\hspace{4.5cm}
$\,{\to}\,\prawy(t{+}1)\,{\to}\,(\rU(m{+}1,t{+}1)\,{\to}\,
\lupa)\,{\to}\,\lupa)$.
\end{itemize}
\noindent First row:
\begin{enumerate}
\item $(\rE(0,0)\to\lupa)\to\start$;
\item $\forall m(\rE(m,0)\,{\to}\, (\rE(m{+}1,0)\,{\to}\,\lupa)
\,{\to}\,\lupa)$.
\end{enumerate}
\noindent First column:
\begin{enumerate}[start=3]
\item
$\forall t(\rE(0,t)\,{\to}\, (\lewy(t{+}1)\,{\to}\, \rE(0,t{+}1)\,{\to}\, 
\lupa)$\\
\nictu\hspace{5cm}
$\,{\to}\,
(\prawy(t{+}1)\,{\to}\, \rE(0,t{+}1)\,{\to}\,\lupa)\,{\to}\,\lupa)$.
\end{enumerate}
\noindent Conclusion:\smallskip
\begin{enumerate}[start=4]
\item $\forall m,t\leq s(\ok(m,t)\,{\to}\,\lupa)$.\smallskip
\end{enumerate}

\noindent Pairs $(m,t)$ should be interpreted as tile locations.
The space to tile is $\NN\times\{0,1\}^*$, not $\NN\times\NN$, so $(m,t)$ 
does not identify a unique location in the tiling, 
but only in the local tiling associated to a certain 
path from $\{0,1\}^*$. An assumption $\rK(m,t)$ 
only states that tile~$\rK$ is to be placed at node $(m,t)$
in the present local tiling. This always refers to some particular
location $(m,w)$, where $|w|=t$.  The predicate $\lewy(t)$ 
(resp.~$\prawy(t)$) indicate that node $w$ is the left (resp.~right) 
child of its parent $w'$, i.e., that $w=w'0$ (resp.~$w=w'1$).

As in Section~\ref{sec:sigmapidwa},
we think of the formulas in~$\Gamma_\G$ as of proof tactics. For 
example, formula $(0_l)$ is used towards the proof goal $\lupa$,
provided $\rK(m,t{+}1),\ldots, \lewy(t{+}1)$ can be verified.
Applying this tactic will not change the proof goal but will 
add $\rT(m{+}1,t{+}1)$ as a~new assumption.

We need to implement the above idea using unary relations and no
arithmetic. In Section~\ref{sec:sigmapidwa} we used different variables 
to represent coordinates of the grid. 
With a fixed supply of variables (cf.~Lemma~\ref{hhfgrtrfg}) we cannot 
do that. However, as long as we only need to encode bounded
values of coordinates, this can be overcome by using
many-argument predicates. 
Those can later be eliminated using Proposition~\ref{lem:overline-to-normal}
to a linear number of unary predicates. 
The basic idea is this. Recall first that the highest
  number which occurs in pairs within~$\Pos(s,s)$ is $2s$.  Assume we
  have two free object variables~$x_0, x_1$, and let us fix a~number
  $n>\log 2s$. \relax Using $x_0$ as 0 and $x_1$ as 1
  one can write a number  $m\leq 2s$ as a~sequence
  $m_1, \ldots, m_n$, where each~$m_i$ is   $x_0$ or $x_1$.  
  A $2n$-ary predicate $\rK(m_1, \ldots, m_n, t_1, \ldots, t_n)$
  can thus be read as~$\rK(m,t)$, where ``variables'' $m$ and $t$ take
  values from $0$ to~\mbox{$2^n\,{-}\,1$}.  This suffices to represent
  coordinates of all points in the set~$\Pos(s,s)$.

The formulas of $\Gamma_\G$ could now be rewritten using $2n$-ary relation 
symbols~$\rT\in\T$ and $n$-ary symbols $\lewy$ and $\prawy$
instead of the binary and unary symbols. In $\rT(m,t)$, the ``variables''
$m$ and~$t$ are now understood as sequences built
from $x_0$ and $x_1$. For instance, $\rE(0,0)$ means 
$\rE(x_0,\ldots,x_0,x_0,\ldots,x_0)$. 
The meaning of a quantifier $\forall m$ is~$\forall m_1m_2\ldots m_n$.

The remaining difficulty is the use of the bound ``${\leq}\,s$'' 
and the successor and 
predecessor operations $+1$ and $-1$. The last two can be handled by
observing that a~number $t'<2^n$ is a successor of a number~$t$
when, for some $k$, the last $k$ bits in the representations of $t'$ 
and~$t$ are, respectively, $011\dots 1$ and $100\dots 0$. 
There are $n$ such patterns (one for each $k$) 
for binary strings of length~$n$. 
Now, for instance, instead of the single formula 

\noindent\hfil 
$\forall m(\rE(m,0)\to (\rE(m+1,0)\to\lupa)\to\lupa)$, 

\noindent 
we can use~$n$ formulas, each following one of those patterns:
\smallskip

\noindent\hfil $\forall\vec z(\rE(\vec z,x_0,\vec{x_1},\vec{x_0})\to 
(\rE(\vec z,x_1,\vec{x_0}, \vec{x_0})\to\lupa)\to\lupa)$.\smallskip

\noindent Above, $\vec z$ is a sequence of $n-k$ bound variables, and 
$\vec{x_1}$ is a sequence of $k-1$ occurrences of~$x_1$.
The symbol $\vec{x_0}$ is loosely used for appropriately long
sequences of $x_0$.

In a similar fashion we can handle the inequality occurring in~(4). 
If $s$ is written in binary as $b_1\ldots b_n$ then a number $m$ can be
less or equal than $s$ in as many different ways as there are numbers~$i<n$ 
with $b_{i+1}=1$. This happens 
when $m$ has the form $b_1\ldots b_i0\vec z$, 
for some~$\vec z$ consisting of $n-i+1$ bits.
So~if $b_{i+1}=b_{j+1}=1$ then we use the formula:\smallskip

\noindent\hfil $\forall\vec z\vec y\,(\ok(b_1\ldots b_i0\vec z,\,
b_1\ldots b_j0\vec y)\to\lupa)$. \smallskip

\noindent
This way we replace each of the assumptions $(0_l)$, $(0_r)$, (1--4),  
by at most $2n^2$ formulas using $2n$-ary predicates.
The set $\Gamma_\G$ consists of all such formulas. To simplify the 
construction in the rest of this section we use the abbreviated notation 
with variables ranging over natural numbers.

\subsection*{Some definitions:}

A finite nonempty set $S\subseteq\Pos(s,s) $ is called 
a~{\it base\/} when 
it is downward closed \wrt~the relation~$\preceq$. 
Then \mbox{$S_2:=\{t\ |\ \exists m.\,(m,t)\in S\}$} is a~finite 
initial segment of~$\NN$. 
A set of formulas $\Niedelta$ is {\it time-coherent for\/}~$S$ iff 
\begin{itemize}
\item for each $t\in S_2$, if $t>0$ then either $\lewy(t)$ or 
$\prawy(t)$ is in~$\Niedelta$,
but not both. 
\end{itemize}
For a time-coherent set $\Niedelta$, we define $a_t^\Niedelta=0$ when 
$\lewy(t)\in\Niedelta$, and $a_t^\Niedelta=1$ otherwise. Denote the
word $a_{1}^{\Niedelta}\ldots a_t^\Niedelta$ by $w_t^\Niedelta$,
and let $w_{0}^{\Niedelta}=\varepsilon$. 
If $t=\max S_2$ then we write $w^\Niedelta$ for $w_t^\Niedelta$. 
We say that $\Niedelta$ is {\it very good for $S$\/} 
when it is time-coherent and the following holds:
\begin{itemize}
\item  Formulas in $\Niedelta$ are only of the forms $\lewy(t)$, $\prawy(t)$,
or $\rT(m,t)$, where $\rT\in\T$.
\item $\rT(m,t)\in\Niedelta$ 
\iff $(m,t)\in S$ and $\TG(m,w_t^\Niedelta)=\rT$.
\item If $\lewy(t)\in \Niedelta$ or $\prawy(t)\in \Niedelta$ then $t\in S_2$.
\end{itemize}

\begin{lem}\label{dsedd1sees}
Let $\Niedelta$ be very good for base $S$.  Assume that 
$\Gamma_\G,\Niedelta\vdash \lupa$. Then $\G$ is $s$-solvable 
from~$w^\Niedelta$.
\end{lem}
\proof  Let $\Gamma_\G,\Niedelta\vdash N:\lupa$, where $N$ is 
a long normal form. We proceed by induction \wrt~$N$, in a similar
style as we did in the proof of Lemma~\ref{tami1}. 

If $N$ begins with a variable of type~$(2)$ then 
$\Gamma_\G,\Niedelta\vdash \rE(m,0)$ (whence the atom $\rE(m,0)$ 
is actually in $\Niedelta$)
and $\Gamma_\G,\Niedelta, E(m+1,0)\vdash\lupa$, for some $m$.
We apply the induction hypothesis to the set $\Niedelta\cup\{E(m+1,0)\}$,
very good for $S\cup\{(m+1,0)\}$. 

In case $N$ begins with a variable of type~$(4)$ we must have 
$\ok(m,t)\in \Niedelta$. In addition, $m,t\leq s$, so the conclusion 
is immediate. The cases ($0_l$) and ($0_r$) are routine as well:
for $\rX=\rT,\rU$, 
we consider the environment $\Niedelta,\rX(m+1,t+1)$, which is very good 
for $S\cup\{(m+1,t+1)\}$. Observe that the location $(m+1,t+1)$ may 
already belong to~$S$, and in this case also the formula $\rX(m+1,t+1)$
is already in $\Niedelta$, because our tiling is deterministic. 

A crucial case is when $N$ begins with an axiom of type (3). 
Then $\Gamma_\G,\Niedelta\vdash \rE(0,t)$, and:
\begin{itemize}
\item $\Gamma_\G,\Niedelta,\lewy(t{+}1),\rE(0,t{+}1)\vdash \lupa$;
\item $\Gamma_\G,\Niedelta,\prawy(t{+}1),\rE(0,t{+}1)\vdash\lupa$.
\end{itemize}
If $t=|w^\Niedelta|$ then the location $(0,t{+}1)$ is ``new'', that is neither
$\lewy(t{+}1)$ nor $\prawy(t{+}1)$ occurs in~$\Niedelta$. Then both 
$\Niedelta,\lewy(t{+}1),\rE(0,t{+}1)$ and 
$\Niedelta,\prawy(t{+}1),\rE(0,t{+}1)$ are very good environments  for 
$S\cup\{(0,t{+}1)\}$. By the induction hypothesis,
 $\G$~is~$s$-solvable from $w^\Niedelta0$ and $w^\Niedelta1$,
so it is $s$-solvable from $w^\Niedelta$ by Lemma~\ref{taksci}.

If $t<|w^\Niedelta|$ then either $\lewy(t{+}1)$ or $\prawy(t{+}1)$ 
is already in $\Niedelta$, and so is $\rE(0,t{+}1)$. Therefore one
of the two environments is identical to $\Gamma_\G,\Niedelta$, and 
we can apply the induction hypothesis (there is a shorter proof). 
\qed

\begin{lem}\label{dsesders11}
If $\Gamma_\G\vdash\start$ then $\G$ is $s$-solvable.
\end{lem}
\proof  A proof of $\start$ is only possible when the judgment 
$\Gamma_\G,\rE(0,0)\vdash\lupa$ is provable. Apply Lemma~\ref{dsedd1sees}.
\relax\qed

\noindent Now we address the question of the converse 
of Lemma~\ref{dsesders11}. 
Put $(m,w)\subseteq (n,v)$ when either $w\subseteq v$ or $w=v$ and $m\leq n$. 
If \mbox{$\TG(m,w)=\ok$}, and $(m,w)$ is minimal \wrt~$\subseteq$, 
then we say that $(m,w)$ is a {\it winning location\/}.

\begin{lem}\label{kkhjhuyg}
Let $\Niedelta$ be very good for base $S$. 
If $\G$ is $s$-solvable from~$w^\Niedelta$ then 
$\Gamma_\G,\Niedelta\vdash \lupa$.
\end{lem}
\proof 
The obvious case is when $\TG(m,w^\Niedelta_r)=\ok$,  for some $(m,r)\in S$,
and $m,r\leq s$. 
Otherwise, for every~$w$ with $|w|=s, w^\Niedelta\subseteq
w$, there is
a winning location $(m,w')$ such that $m\leq s$ and $w'\subseteq w$.
(Some of these winning locations may be 
equal, in particular if $w'\subseteq w^\Niedelta$ then all of them are equal.)
For every such $(m,w')$, the {\it distance\/} from $S$ to $(m,w')$
is the cardinality of the difference $\Pos(m,|w'|)-S$.

The proof of the 
lemma is by induction \wrt~the sum of distances from~$S$ to all 
winning locations $(m,w')$ such that $w'\subseteq w^\Niedelta$ or 
$w^\Niedelta\subseteq w'$.
(This is equivalent to saying that there is 
$w$ of length~$s$ such that $w',w^\Niedelta\subseteq w$.)

The base case has already been treated, so assume that 
we have a winning location $(m,w')$. 
If $w'\subseteq w^\Niedelta$ then 
it is the only winning location of interest. 
If~$(m, |w'|)\not\in S$ then the difference
 $\Pos(m, |w'|)-S$ has a~minimal element 
$(m_1, t_1)$. As in the proof of Lemma~\ref{nazad},
we apply the induction hypothesis to the set $S\cup\{(m_1, t_1)\}$
and apply an assumption~$(0_l)$ or~$(0_r)$. 
This we can do because either $\lewy(t_1)$ or $\prawy(t_1)$ must
belong to~$\Niedelta$. 

The remaining case is when 
there are at least two winning locations, all
of them of the form $(m,w^\Niedelta_v)$
with $v\neq\varepsilon$. In addition,
at least one such $v$ begins with 0 and at least one with~1. 
Take $t=|w^\Niedelta|$ and let  
$S'=S\cup\{(0,t+1)\}$; then the environments 
\mbox{$\Niedelta_0=\Niedelta\cup\{\lewy(t{+}1),\rE(0,t{+}1)\}$}
and $\Niedelta_1=\Niedelta\cup\{\prawy(t{+}1),\rE(0,t{+}1)\}$
are very good for $S'$, so $\Gamma_\G,\Niedelta_0\vdash\lupa$
and $\Gamma_\G,\Niedelta_1\vdash\lupa$ by the induction hypothesis.
(In each case, we have fewer winning positions and thus fewer
components in our sum.) We now use asumption (3). 
\relax\qed

\begin{lem}\label{fdfdre22ess}
If $\G$ is $s$-solvable then $\Gamma_\G\vdash\start$.
\end{lem}
\proof  Apply Lemma~\ref{kkhjhuyg} to the base $S=\pusty$.
\relax\qed

\begin{thm}\label{ggfwasdee}
The decision problem for monadic
$\Sigma_1$ is co-{\sc Nexptime}-hard. 
\end{thm}
\proof  
Hardness for arbitrary predicates follows from Lemmas~\ref{unavoiis} 
and~\ref{fdfdre22ess}. Translation to the monadic case is possible 
by Proposition~\ref{lem:overline-to-normal} 
because all formulas we use are easy. 
\qed

In fact our hardness result applies to a ``shallow'' fragment of  monadic
$\Sigma_1$. This fragment consists of formulas of the form
$\tau_1\to\cdots\to\tau_n\to{\bf a}$, where $\tau_i$ are universal
formulas, i.e., we have $\tau_i=\forall\vec x_i\, \tau_i'$ with 
quantifier-free~$\tau_i'\,$.

\subsection{Arity-bounded refutation soup}

For the matching upper bound we show that provability of 
arity-bounded $\Sigma_1$ formulas is solvable in {\it co}-{\sc Nexptime}. 
More precisely, let us fix a~number $r$ and consider only formulas 
involving predicates of arity at most~$r$. 
We will demonstrate a nondeterministic exponential time algorithm 
for non-provability, i.e., refutability of such formulas. 

\begin{lem}\label{piekne}~
\begin{enumerate}\parskip=0pt
\item Every $\Sigma_1$ formula has the form 
$\tau_1\to\tau_2\to\cdots\to\tau_n\to{\bf a}$, where $\tau_i\in\Pi_1$,
and $\,{\bf a}$ is an atom.
\item Every $\Pi_1$ formula has the form
$\forall\vec y_1(\sigma_1\to \forall\vec y_2(\sigma_2\to 
\cdots\to\forall\vec y_k(\sigma_k\to{\bf a})\ldots))$, where 
$\sigma_i\in \Sigma_1$, and $\,{\bf a}$ is an atom.
\end{enumerate}
\end{lem}
\proof  The pseudo-grammar of Section~\ref{theminc} simplifies 
as follows. The 
metavariable~{\bf a} now stands for an atom of arity $r$ or less.
\begin{itemize}\parskip=0pt
\item $\Sigma_{1} ::= {\bf a}\ |\ \Pi_{1}\to \Sigma_{1}$; 
\item $\Pi_{1} ::= {\bf a}\ |\ \Sigma_{1}\to\Pi_{1}\ |\ 
\forall x\,\Pi_{1}\,$.\hfill\qed
\end{itemize}
As it suffices to deal with long normal proofs, we are mostly interested 
in judgments  of the form $\Gamma\vdash N:{\bf a}$, where $\Gamma$ 
consists of $\Pi_1$ formulas, and ${\bf a}$ is an atom.
The long normal proof $N$ must begin with a proof variable $X$; assume that
a declaration of the form 
\mbox{$X: \forall\vec y_1(\sigma_1\to \forall\vec y_2(\sigma_2\to 
\cdots\to\forall\vec y_k(\sigma_k\to{\bf b})\ldots))$} is in $\Gamma$.
Types $\sigma_i$ are in $\Sigma_1$ and therefore 
$\sigma_j=\tau_{j1}\to\tau_{j2}\to\cdots\to\tau_{jr_j}\to{\bf a}_j$,
for $j=1,\ldots, k$.
To maintain some basic hygiene we assume that all variables
in $\vec y_1\vec y_2\dots\vec y_k$ are different and not free in~$\Gamma$. 
Then we have $N=X\vec x_1(\lambda Y_{11}\ldots Y_{1r_1}.\, N_1)\ldots 
\vec x_k(\lambda 
Y_{k1}\ldots Y_{kr_k}.\, N_k)$, for some variables 
$\vec x_1\vec x_2\ldots\vec x_k$ (assumed different from~$\vec y_i$). 
Write $S$ for the substitution 
$[\vec y_1:=\vec x_1,\vec y_2:=\vec x_2,\ldots,\vec y_k:=\vec x_k]$.
Then we must have $\,{\bf a}={\bf b}[S]$, and 
$\,\Gamma,Y_{j1}:\tau_{j1}[S],Y_{j2}:\tau_{j2}[S],\ldots,Y_{jr_j}:
\tau_{j_n}[S]\vdash N_j:{\bf a}_j[S]$, for all~$j$.

We know from Lemma~\ref{hhfgrtrfg} 
that if $\Gamma\vdash\varphi$ is inhabited,
then there exists a long normal inhabitant~$N$ with 
$\FV(N)\subseteq \FV(\Gamma)\cup\FV(\varphi)$. 
Therefore, the above analysis can be strenghtened by the 
requirement that all the variables $\vec x_1\vec x_2\ldots\vec x_k$ 
are in $\FV(\Gamma)\cup\FV({\bf a})$.
The next lemma is a contraposition of the above 
taking this additional requirement into account.

\begin{lem}\label{wkazdym}Let $\Gamma$ consist of $\Pi_1$ formulas 
and let ${\bf a}$ be an atom. Then $\Gamma\nvdash {\bf a}$ \iff
\begin{itemize}\parskip=0pt
\item[~] for every $X\in\dom\Gamma$ of type 
$\forall\vec y_1(\sigma_1\to \forall\vec y_2(\sigma_2\to 
\cdots\to\forall\vec y_k(\sigma_k\to{\bf b})\ldots))$,
\item[~]  every $S$ with $\dom S=\vec y_1,\ldots,\vec y_n$,
${\rm Rg}(S)\subseteq \FV(\Gamma)\cup\FV({\bf a})$,
and $\,{\bf a}={\bf b}[S]$,
\item[~] there is $j\in\{1,\ldots,k\}$ with 
$\sigma_{j}=\tau_{j1}\to\tau_{j2}\to\cdots\to\tau_{jr_j}\to{\bf a}_j$
\item[~] such that  
$\,\Gamma,\tau_{j1}[S],\tau_{j2}[S],\ldots,\tau_{jr_j}[S]\nvdash {\bf a}_j[S]$.\qed
\end{itemize}
\end{lem}
\noindent 
Morally, Lemma~\ref{wkazdym} states that in a certain
proof-construction game one of the players has a~winning strategy: either
the Prover, trying to construct a long normal proof (always a~finite one) 
or the Reviewer, attempting to build a (possibly infinite) 
{\it refutation\/}, cf.~\cite{skura-handbook}. Indeed, 
let $\Gamma\vdash {\bf a}$ be a $\Sigma_1$ judgment as above
and assume the notation from Lemma~\ref{wkazdym}.
\mbox{Every} pair $(X,S)$ such that a declaration 
\mbox{$X:\forall\vec y_1(\sigma_1\to \forall\vec y_2(\sigma_2\to 
\cdots\to\forall\vec y_k(\sigma_k\to{\bf b})\ldots))$} is in $\Gamma$,
and $S$ is a~variable substitution satisfying $\,{\bf a}={\bf b}[S]$,
 is called a {\it question induced by\/} $\Gamma\vdash {\bf a}$. 
For any~\mbox{$j\in\{1,\ldots,k\}$}, a~{\it $j$-th answer\/} to the 
question $(X,S)$ is any judgment
$\,\Gamma',\tau_{j1}[S],\tau_{j2}[S],\ldots,\tau_{jr_j}[S]\vdash {\bf a}_j[S]$,
where $\Gamma\subseteq\Gamma'$. (Note that we do not require
$\Gamma=\Gamma'$. This extra flexibility is used in the proof of 
Lemma~\ref{czynsze}.)

Lemma~\ref{wkazdym} may now be read as: $\Gamma\nvdash {\bf a}$ 
\iff~Reviewer can answer every question (and in addition 
$\Gamma'=\Gamma$ always holds). 
This constitutes a refutation seen as a Reviewer's winning strategy. 
A compact way to represent such a refutation is simply a set of judgments.

 A {\it refutation soup\/} 
is a non-empty set $\Z$ of $\Sigma_1$ judgments such~that 
\begin{itemize}\parskip=0pt
\item If $\Gamma\vdash{\bf a}$ is in $\Z$ then for every question 
induced by $\Gamma\vdash{\bf a}$ there is an answer in~$\Z$. 
\end{itemize}
We say that $\Z$ {\it refutes\/} $\Gamma_0\vdash{\bf a}_0$ whenever 
the judgment $\Gamma_0\vdash{\bf a}_0$ belongs to~$\Z$. Then we also say that
$\Gamma_0\vdash{\bf a}_0$ is {\it refutable\/}. 
Observe that a~judgment of the form $\Gamma,X:{\bf a}\,\vdash {\bf a}$
cannot occur in a soup. Indeed, there is no answer to the question 
$\<X,\pusty\>$. 

\begin{lem}\label{coza}
A judgment $\Gamma_0\vdash{\bf a}_0$ is refutable \iff 
$\Gamma_0\nvdash{\bf a}_0$.
\end{lem}
\proof  $(\To)$ Let $\Z$ be a refutation soup such that
some judgments in $\Z$ are provable. Among such judgments there
is one which has a~shortest long normal
proof. Let $\Gamma_0\vdash {\bf a}_0$  be this judgment.
Assume that $N=X\vec x_1(\lambda Y_{11}\ldots Y^1_{1r_1}.\, N_1)\ldots 
\vec x_k(\lambda Y_{k1}\ldots Y_{kr_k}.\, N_k)$ is 
the proof. Note that $k\neq 0$, as otherwise the question 
$(X,\pusty)$ has no answer. Consider the question~$(X,S)$, where 
\mbox{$S=[\vec y_1:=\vec x_1,\vec y_2:=\vec x_2,\ldots,\vec y_k:=\vec x_k]$}. 
In the refutation~$\Z$ there is an answer of the form 
$\,\Gamma',\tau_{j1}[S],\tau_{j2}[S],\ldots,\tau_{jr_j}[S]\vdash {\bf a}_j[S]$.
Clearly, $\Z$ refutes this judgment as well. But on the other hand,
$\,\Gamma',\tau_{j1}[S],\tau_{j2}[S],\ldots,\tau_{jr_j}[S]\vdash N_j 
:{\bf a}_j[S]$, i.e., the refuted judgment has a~proof, shorter than~$N$.
This contradicts our assumption about~$N$.

$(\Ot)$ A soup may be defined as a sum of an ascending sequence 
of sets $\Z_n$. The set~$\Z_0$ consists only of the initial 
judgment $\Gamma_0\vdash{\bf a}_0$. Then, for any $n$,
we select a judgment in~$\Z_{n}$ and a~question induced by this judgment 
which does not have an answer in $\Z_{n}$. By Lemma~\ref{wkazdym}
there is always a non-provable answer. We obtain $\Z_{n+1}$ 
by adding this answer to $\Z_n$. This process must end because 
only a finite number of judgments may occur in the construction.
\qed

We now show that every refutable judgment has a small soup.

\begin{lem}\label{czynsze}
If $\Gamma_0\nvdash{\bf a}_0$ then there is a~refutation soup of size 
exponential in the length $n$ of the judgment $\Gamma_0\vdash{\bf a}_0$.
\end{lem}
\proof  
Let $\F$ be the set of all formulas of the form $\tau[S]$, where 
$\tau$ is a subformula of a formula in $\Gamma_0$, 
and $S$ is a substitution 
such that ${\rm Rg}(S)\subseteq \FV(\Gamma_0)\cup\FV({\bf a}_0)$.
A judgment $\Gamma\vdash {\bf a}$ is {\it reasonable\/} when
$\Gamma\cup\{{\bf a}\}\subseteq \F$. 

As in the proof of Lemma~\ref{coza} we construct a soup by induction,
but now we introduce some structure: rather than just a set 
of judgments we define a tree labeled by judgments. The rule is that 
children of every node are answers to questions induced by that node. 
We begin from the root labeled $\Gamma_0\vdash{\bf a}_0$. At every step 
we select a leaf node $\Gamma\vdash {\bf a}$ 
(a judgment not processed before) and for every 
question $(X,S)$  
induced by that node we choose an unprovable 
answer to that question, say $\Gamma'\vdash {\bf a}'$,
which is reasonable and {\it maximal\/} in the following 
sense:~whenever $\Gamma'\subnoteq \Gamma''\subseteq\F$ 
then $\,\Gamma''\vdash  {\bf a}_j[S]$ has a proof.  
Then we add $\Gamma'\vdash {\bf a}'$ as a new child of $\Gamma\vdash {\bf a}$,
unless $\Gamma'\vdash {\bf a}'$ already occurs on the path from the 
root to $\Gamma\vdash {\bf a}$.
It~should be clear that the set of all labels in our tree 
is a soup.

If a non-root judgment $\Gamma\vdash {\bf a}$ is an ancestor 
of $\Gamma'\vdash {\bf b}$ in our tree then $\Gamma\subseteq\Gamma'$.
Therefore ${\bf a}\neq {\bf b}$, as otherwise 
$\Gamma\vdash {\bf a}$ would not be selected as maximal, or the same 
judgment would occur twice on a path. 
It follows that every path of the tree is of length at most $n^{r+1}$,
where~$r$ is the maximum arity of predicates in $\F$.
Indeed, every judgment in a non-root position
along the path addresses a different target,
and there is at most $n\cdot n^r$ of those (up to $n$ predicates times up to 
$n^r$ ways in which $n$ variables can occur at $r$ positions). 
Since the maximal branching is $n\cdot n^n\cdot n^n$ (an upper bound for 
the number of questions), the total number of nodes does not 
exceed~$(n\cdot n^n\cdot n^n)^{n^{r+1}}\leq   2^{n^{r+4}}$. 
\qed

\begin{proposition} For every $r$, non-provability of  
$\Sigma_1$ formulas using at most $r$-ary predicates 
is solvable in {\sc Nexptime}.
\end{proposition}
\proof A nondeterministic algorithm can generate a refutation soup
and verify its correctness in exponential time. 
\qed

In particular we have:

\begin{cor}\label{cddgerfe} The decision problem for 
$\Sigma_1$ formulas of any fixed finite signature 
is in the class {\it co}-{\sc Nexptime}.\qed
\end{cor}

Together with Theorem~\ref{ggfwasdee} we obtain the final result.

\begin{thm} 
\label{thm:co-nexptime-complete}
For every $r \in\NN$, the decision problem for 
$\Sigma_1$ formulas using at most $r$-ary predicates is 
{\it co}-{\sc Nexptime}-complete.\qed
\end{thm}

\section{Conclusion and future work}
\label{sec:conclusions}

We proved that derivability of universally-implicational 
formulas for the class $\Delta_2$ of Mints hierarchy (and therefore
for all larger classes) is undecidable even for unary predicate symbols.
In case of $\Sigma_1$ the problem is in general {\sc Expspace}-complete,
but it turns out only {\it co}-{\sc Nexptime}-complete if we restrict the 
arity of predicates (this applies e.g., to the monadic fragment). 
In particular the exponential upper bound holds for every finite 
signature.

These results combined with an earlier analysis \cite{suw2013}
give the picture of complexity of provability in Mints hierarchy in
which the level of a~formula~$\varphi$ is determined by the level of a~prenex 
formula classically equivalent to $\varphi$. Observe that all the hardness 
results were obtained for formulas with a fixed depth
of quantifiers. 

The fragment of intuitionistic logic discussed in this paper 
only involves the two basic connectives, $\forall$ and $\to$.
By conservativity, all our lower bounds extend to the full 
first-order language with $\exists$, $\vee$, $\wedge$, and~$\bot$. 
It is not necessarily so with the upper bounds. The exponential
space algorithm for $\Sigma_1$ extends to the general case, but
the refutation soup argument does not (because targets in judgments
can be disjunctions). We conjecture that the $\Sigma_1$ fragment 
of the full first-order     
logic will turn out {\sc Expspace}-complete even in the monadic case. 
On the other hand, we believe that the number of predicates matters:
perhaps Corollary~\ref{cddgerfe} can be improved down to
{\sc Pspace}? 

Another issue demanding future work is the exact complexity of 
the class~$\Pi_1$~\cite{suw2013-www,suw2013}.

\bibliographystyle{plain}
\bibliography{minc} 

\begin{thebibliography}{10}

\bibitem{Bonevac}
Daniel Bonevac.
\newblock A history of quantification.
\newblock In {\em Logic: A History of its Central Concepts}, volume~11 of {\em
  Handbook of the History of Logic}. North Holland, 2012.

\bibitem{BorgerGG1997}
Egon B{\"o}rger, Erich Gr{\"a}del, and Yuri Gurevich.
\newblock {\em The Classical Decision Problem}.
\newblock Perspectives in Mathematical Logic. Springer, 1997.

\bibitem{Agda}
Ana Bove, Peter Dybjer, and Ulf Norell.
\newblock A brief overview of {A}gda -- a functional language with dependent
  types.
\newblock In {\em Theorem Proving in Higher Order Logics}, volume 5674 of {\em
  LNCS}, pages 73--78. Springer, 2009.

\bibitem{Burr09}
Wolfgang Burr.
\newblock The intuitionistic arithmetical hierarchy.
\newblock In {\em Logic Colloquium '99}, volume~17 of {\em Lecture Notes in
  Logic}, pages 51--59. ASL, 1999.

\bibitem{church-iml}
Alonzo Church.
\newblock {\em Introduction to Mathematical Logic}.
\newblock Princeton, 1944.

\bibitem{Coq}
{Coq Development Team}.
\newblock {\em The {Coq} Proof Assistant Reference Manual {V8}.4}, March 2012.
\newblock \url{http://coq.inria.fr/distrib/V8.4/refman/}.

\bibitem{DegtyarevGNVV00}
Anatoli Degtyarev, Yuri Gurevich, Paliath Narendran, Margus Veanes, and Andrei
  Voronkov.
\newblock Decidability and complexity of simultaneous rigid {E}-unification
  with one variable and related results.
\newblock {\em Theoretical Computer Science}, 243(1-2):167--184, 2000.

\bibitem{dowekjiang06}
Gilles Dowek and Ying Jiang.
\newblock Eigenvariables, bracketing and the decidability of positive minimal
  predicate logic.
\newblock {\em Theoret. Comput. Sci.}, 360(1--3):193--208, 2006.

\bibitem{Fitting81}
M.~Fitting.
\newblock {\em Fundamentals of Generalized Recursion Theory}.
\newblock Elsevier, 1981.

\bibitem{Fleischmann10}
Jonathan Fleischmann.
\newblock Syntactic preservation theorems for intuitionistic predicate logic.
\newblock {\em Notre Dame Journal of Formal Logic}, 51(2):225--245, 2010.

\bibitem{Immerman99}
Neil Immerman.
\newblock {\em Descriptive Complexity}.
\newblock Springer, 1999.

\bibitem{Kreisel58}
G.~Kreisel.
\newblock Elementary completeness properties of intuitionistic logic with a
  note on negations of prenex formulae.
\newblock {\em J. Symbolic Logic}, 23(3):pp. 317--330, 1958.

\bibitem{Kusmierek07}
Dariusz Ku\'smierek.
\newblock The inhabitation problem for rank two intersection types.
\newblock In {\em TLCA}, volume 4583 of {\em LNCS}, pages 240--254. Springer,
  2007.

\bibitem{minc}
G.E. Mints.
\newblock Solvability of the problem of deducibility in {LJ} for a class of
  formulas not containing negative occurrences of quantifiers.
\newblock {\em Steklov Inst.}, 98:135--145, 1968.

\bibitem{orewkow65}
V.P. Orevkov.
\newblock The undecidability in the constructive predicate calculus of the
  class of formulas of the form $\neg\neg\forall\exists$.
\newblock {\em Doklady AN SSSR}, 163(3):581--583, 1965.

\bibitem{orewkow76}
V.P. Orevkov.
\newblock Solvable classes of pseudoprenex formulas.
\newblock {\em Zapiski nauchnyh Seminarov LOMI}, 60:109--170, 1976.

\bibitem{Paulson89}
Lawrence~C. Paulson.
\newblock The foundation of a generic theorem prover.
\newblock {\em Journal of Automated Reasoning}, 5(3):363--397, 1989.

\bibitem{rash54}
H.~Rasiowa and R.~Sikorski.
\newblock On existential theorems in non-classical functional calculi.
\newblock {\em Fundamenta Mathematicae}, 41:21--28, 1954.

\bibitem{RehofU12}
Jakob Rehof and Paweł Urzyczyn.
\newblock The complexity of inhabitation with explicit intersection.
\newblock In {\em Logic and Program Semantics}, volume 7230 of {\em LNCS},
  pages 256--270. Springer, 2012.

\bibitem{rosen05}
Eric Rosen.
\newblock On the first-order prefix hierarchy.
\newblock {\em Notre Dame Journal of Formal Logic}, 46(2):147--164, 2005.

\bibitem{rummelhoff}
Ivar Rummelhoff.
\newblock {\em Polymorphic ${\Pi} 1$ Types and a Simple Approach to
  Propositions, Types and Sets}.
\newblock PhD thesis, University of Oslo, 2007.

\bibitem{suw2013-www}
Aleksy Schubert, Paweł Urzyczyn, and Daria Walukiewicz-Chrząszcz.
\newblock Restricted positive quantification is not elementary.
\newblock In Hugo Herbelin, Pierre Letouzey, and Matthieu Sozeau, editors, {\em
  Proc.~TYPES 2014}, volume~39 of {\em LIPIcs}, pages 251--273. Schloss
  Dagstuhl--Leibniz-Zentrum f\"ur Informatik, 2015.

\bibitem{suw2013}
Aleksy Schubert, Paweł Urzyczyn, and Daria Walukiewicz-Chrząszcz.
\newblock How hard is positive quantification?
\newblock To appear in {\it ACM ToPLaS}, 2016.

\bibitem{suz-fos15}
Aleksy Schubert, Paweł Urzyczyn, and Konrad Zdanowski.
\newblock On the {M}ints hierarchy in first-order intuitionistic logic.
\newblock In A.~Pitts, editor, {\em Foundations of Software Science and
  Computation Structures 2015}, volume 9034 of {\em Lecture Notes in Computer
  Science}, pages 451--465. Springer, 2015.

\bibitem{skura-handbook}
Tomasz Skura.
\newblock Refutation systems in propositional logic.
\newblock In Dov~M. Gabbay and Franz Guenthner, editors, {\em Handbook of
  Philosophical Logic}, volume~16, pages 115--157. Springer, second edition,
  2011.

\bibitem{sorm06}
M.H. S{\o}rensen and P.~Urzyczyn.
\newblock {\em Lectures on the Curry-Howard Isomorphism}, volume 149.
\newblock Elsevier, 2006.

\bibitem{urzy09}
P.~Urzyczyn.
\newblock Inhabitation of low-rank intersection types.
\newblock In P.-L. Curien, editor, {\em TLCA}, volume 5608 of {\em LNCS}, pages
  356--370. Springer, 2009.

\bibitem{Wang60}
Hao Wang.
\newblock Toward mechanical mathematics.
\newblock {\em IBM J. Res. Dev.}, 4(1):2--22, January 1960.

\end{thebibliography}
\end{document}